\documentclass[12pt]{article}
\usepackage{graphicx}
\usepackage[round]{natbib}
\usepackage{url} 

\usepackage{hyperref}

\usepackage{adjustbox}
\usepackage[utf8]{inputenc}
\usepackage{amsmath,amsthm,amsfonts,amssymb}
\usepackage{xcolor}
\usepackage{nomencl}
\usepackage{soul}
\usepackage{subfigure}
\usepackage[center]{caption}
\usepackage{graphicx}
\usepackage{multirow}
\usepackage{array}

\makeatletter

\providecommand{\tabularnewline}{\\}

\newtheorem{theorem}{Theorem}
\newtheorem{prop}[theorem]{Proposition}
\newtheorem{corollary}[theorem]{Corollary}

\newcommand{\blind}{1}

\usepackage[colorinlistoftodos]{todonotes}

\begin{document}

\date{}
\def\spacingset#1{\renewcommand{\baselinestretch}%
{#1}\small\normalsize} \spacingset{1}

\title{Adaptive Partially-Observed Sequential Change Detection and Isolation    }
\if1\blind
{
  \author{Xinyu Zhao$^1$, Jiuyun Hu$^1$, Yajun Mei$^2$, Hao Yan$^1$
  \thanks{
    The authors gratefully acknowledge the support from \textit{NSF DMS 1830363, and CMMI 1922739}}
    \\$^1$School of Computing and Augmented Intelligence \\
    Arizona State University, \\
    $^2$School of Industrial and Systems Engineering, \\ 
    Georgia Institute of Technology \\
}
  
} \fi
\maketitle

\bigskip
\begin{abstract}
High-dimensional data has become popular due to the easy accessibility of sensors in modern industrial applications. However, one specific challenge is that it is often not easy to obtain complete measurements due to limited sensing powers and resource constraints. Furthermore, distinct failure patterns may exist in the systems, and it is necessary to identify the true failure pattern. This work focuses on the online adaptive monitoring of high-dimensional data in resource-constrained environments with multiple potential failure modes. To achieve this, we propose to apply the Shiryaev–Roberts procedure on the failure mode level and utilize the multi-arm bandit to balance the exploration and exploitation. We further discuss the theoretical property of the proposed algorithm to show that the proposed method can correctly isolate the failure mode. Finally, extensive simulations and two case studies demonstrate that the change point detection performance and the failure mode isolation accuracy can be greatly improved.
\end{abstract}

\noindent%


\spacingset{2} 

\section{Introduction}
Nowadays, most industrial applications are instrumented with hundreds or thousands of sensors due to the advancement in sensing technology.  Real-time process monitoring and fault diagnosis are among the benefits that can be gained from effective modeling and analysis of the produced high-dimensional streaming data. Classical researches for process monitoring of high-dimensional streaming data focus on a fully observable process, which means at each sampling time point, all the variables can be observed for analysis \citep{yanRealTime2018}. 
However, it is often infeasible to acquire measurements of all these sensing variables in real time due to limited sensing resources, sensing capacity, sensor battery, or other constraints such as system transmission bandwidth, memory, storage space, and processing speed in modern industrial applications  \citep{liuAdaptive2015}. 
Furthermore, under change detection and isolation setting, we assume that the engineered systems that are being studied have several distinct failure modes and patterns but do not know which failure mode may occur beforehand. Overall, this paper focuses on change point detection under resource-constrained environments with multiple potential failure modes.

\begin{figure}
    \centering
    \subfigure[Hot foaming process]{\includegraphics[width=0.32\textwidth]{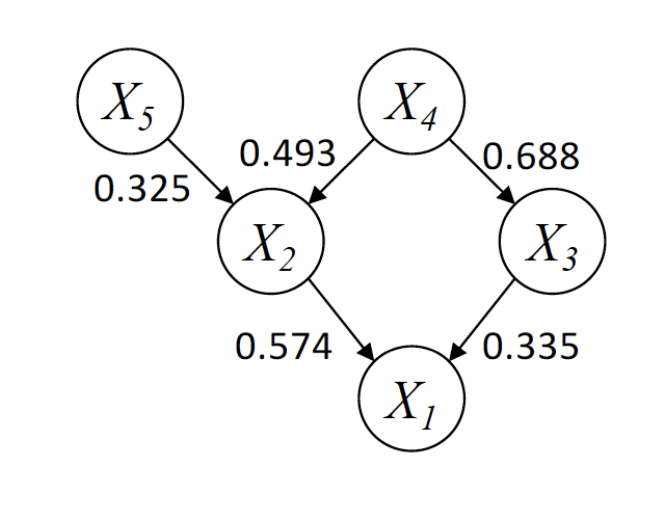}\label{fig:hot}}
    \subfigure[3D printing example]{\includegraphics[width=0.32\textwidth]{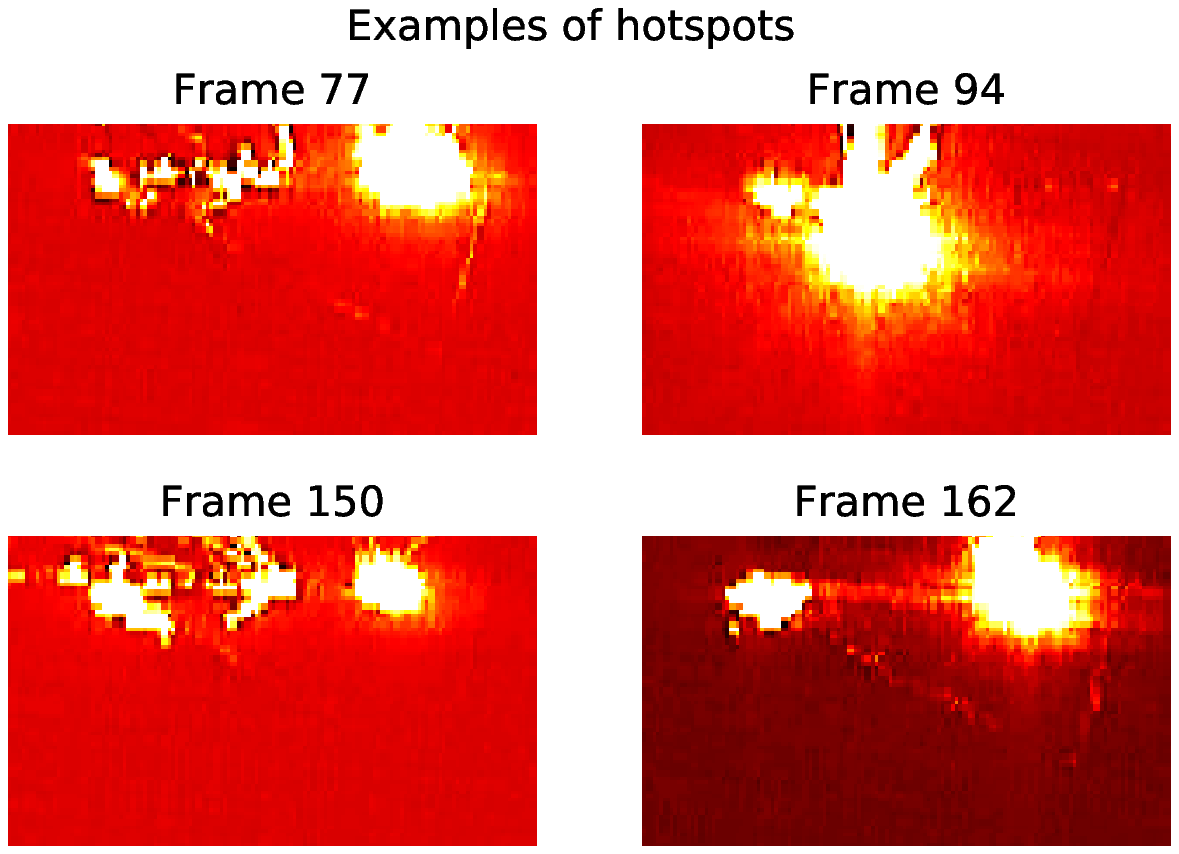} \label{fig:3d_example}}
    
    \caption{Examples of Complex data in Various Industrial Applications \textbf{Left} figure shows an example of a hot foaming process.  \textbf{Right} figure shows an example of monitoring the thermal images in additive manufacturing. }
    \label{fig:casestudy}
    
\end{figure}

The first motivating example is in the hot forming process \citep{liOptimal2010} as shown in Fig.~\ref{fig:hot}. There are five sensing variables in the system: the final dimension of workpiece ${\bf X}_1$, the tension in workpiece ${\bf X}_2$, material flow stress ${\bf X}_3$, temperature ${\bf X}_4$, and blank holding force ${\bf X}_5$. These five variables can be represented as a Bayesian network, as shown in Fig.~\ref{fig:hot}. For example, if we know that the change of ${\bf X}_4$ and ${\bf X}_5$ are the two major failure sources in the system. If ${\bf X}_4$ changes,  $({\bf X}_1,{\bf X}_2, {\bf X}_3, {\bf X}_4)$ will also change. Furthermore, If ${\bf X}_5$ changes, only $({\bf X}_1, {\bf X}_2, {\bf X}_5)$ will change. Therefore, different failure modes may affect a different subset of sensors differently. 


Another example comes from in-situ hot-spots detection in the laser powder bed fusion (LPBF) process in the metal additive manufacturing process. A thermal camera is often used to monitor the stability of the process while the product is being produced on a layer-by-layer basis. Here, detecting the hot-spots early is crucial for further product quality control. Fig.~\ref{fig:3d_example} show an example of such hot-spots from the thermal camera. Given that the anomaly or hot-spots can only occur on the edge/corner of the scanning path, multiple failure modes can be defined.

\begin{figure} 
    \centering
    \vspace*{-0.2cm}
    \includegraphics[width=0.6\textwidth]{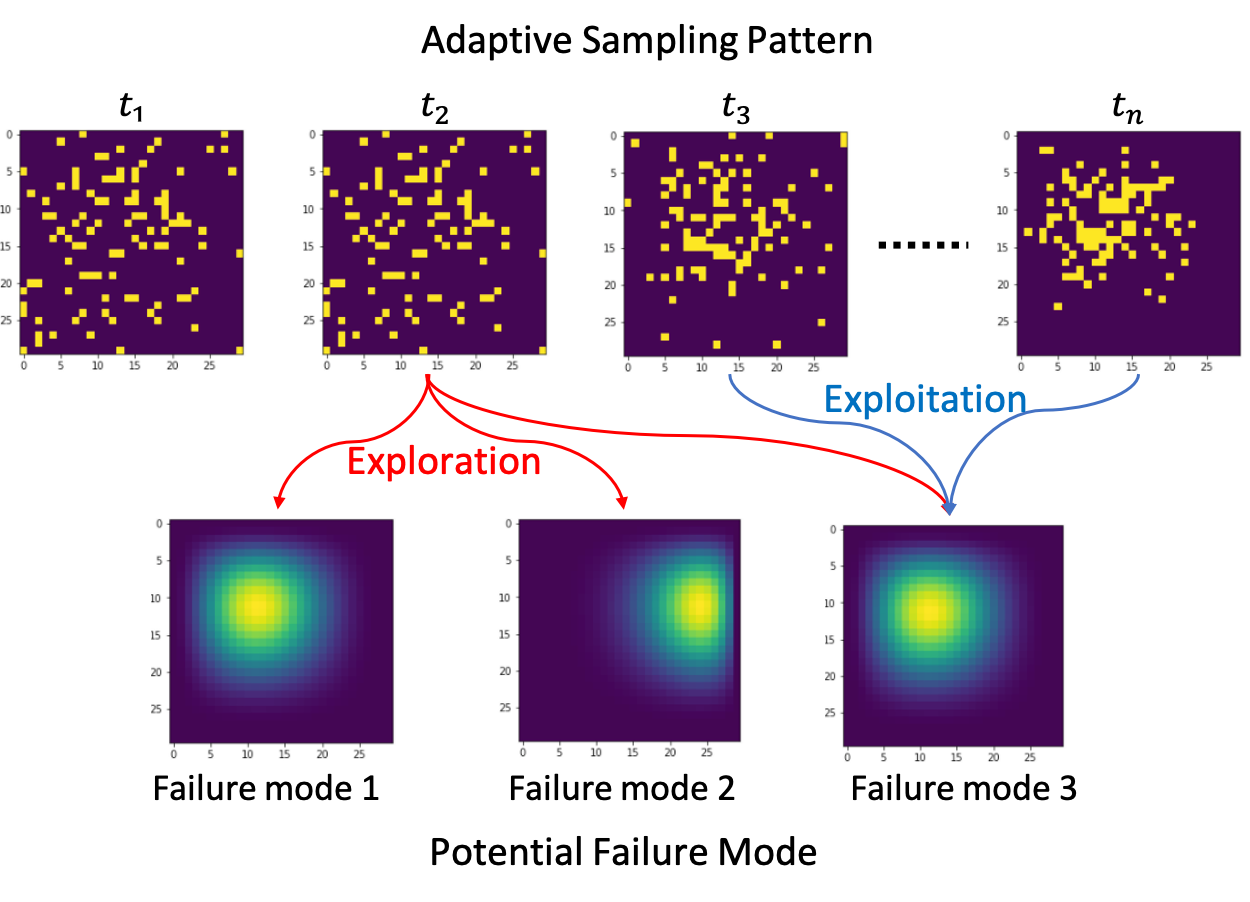}
    \caption{Conceptual Illustration of the Balance of Exploration and Exploitation; The sampling patterns at $t_1, t_2$ focus on the exploration of all failure modes. The sampling pattern at $t_3,\cdots,t_n$ focuses on the exploitation of the failure mode 3. }
    \label{fig:eeblance}
\end{figure}

There are a few challenges of sequential change-point detection under the sampling constraint: 1) From the previous examples, the failure mode distribution can be quite complicated. For example, in the hot foaming process, as shown in Fig.~\ref{fig:hot}, we aim to detect the failure mode with the weakly conditional dependency on the graph; In the laser powder bed fusion process, as shown in Fig.~\ref{fig:3d_example}, we aim to detect the spatially clustered hot-spots. 2) Even though we assume that we have prior knowledge of different potential failure modes, we do not know which failure mode may occur in the system. The main challenge is to balance the exploration of all potential failure modes and the exploitation to focus on the most probable failure mode. A conceptual illustration of the proposed algorithm is provided in Fig.~\ref{fig:eeblance}. The illustration example has shown an example that the sampled points are performed on the 2D spatial domain. The sampling patterns at time $t_1$ and $t_2$ focus on exploration for all failure modes and the sampling patterns at $t_3,\cdots,t_n$ focus on exploitation for failure mode 3. In general, it is hard to decide when the algorithms should change to exploitation or which failure mode they should focus on. 
Finally, given that the multiple failure modes have quite complex shapes and distributions, the exploration and exploitation among these modes are often quite challenging. 

There are also many works focusing on change-point detection under resources constraint. Most of the existing works are proposed based on the "local monitoring and global decision" framework, which focuses on monitoring each data stream independently using local monitoring statistics and then fusing these local monitoring statistics together via a global decision framework. For example, \cite{liuAdaptive2015} proposed scalable and efficient algorithms for adaptive sampling for online monitoring. The method introduced a compensation parameter for the unobserved variables to increase the chance of exploring them. Recently, \cite{zhang2020bandit} proposed to combine the powerful tools of the multi-arm bandit problem for efficient real-time monitoring of HD streaming data. However, these works either assume the data stream is independent or cannot take advantage of the failure mode information in some systems, which fails to monitor and identify the correct failure pattern. For a complete literature review of monitoring of high-dimensional streaming data, please see Section \ref{sec: review}. 

To generalize the sequential change-point detection framework to both detect and identify the correct failure modes, \textit{change detection and isolation literature} has been proposed in the literature, which also inspires this research. The change-point detection and isolation often assume that there are a set of pre-defined post-change distributions. The goal is not only to detect the change with the shortest detection delay but also to identify which change mode occurs in the system. For example, \cite{chen2020bayesian} proposed a Bayesian method to decide on a procedure to identify both the change point as well as the correct change mode. 
For a complete literature review of change detection and isolation, please see Section 3.3. However, these works typically assume that the data is fully observed, which cannot be applied to partially observed data.

To address the challenge of multiple failure modes and partially observed data, we propose a novel Multiple Thompson Sampling Shiryaev-Roberts-Pollak (MTSSRP) Method by a modified "local monitoring and global decision framework". As far as the authors know, this is the first work that discusses the adaptive sampling framework for failure mode detection and isolation. Unlike the literature on the monitoring of HD streaming data, where the local monitoring statistics are defined at each individual sensor, we propose to define the local statistics for each individual failure mode. This enables the proposed MTSSRP to take advantage of the failure mode information, which is very important in the high-dimensional space, given that failure mode information can significantly reduce the search space since there are unlimited ways that change may occur in the high-dimensional space. To quantify the uncertainty of unobserved sensing variables for different failure modes, we propose to apply the Shiryaev-Robert (SR) procedure for sequential change point detection on the failure mode level.  

Furthermore, to balance the exploration and exploitation, we will borrow the idea from Multi-arm Bandit (MAB). MAB aims to sequentially allocate a limited set of resources between competing "arms" to maximize their expected gain, where the reward function for each arm is not known. MAB provides a principled way to balance exploration and exploitation. To apply MAB for change point detection under the sampling constraint, we propose to use the SR statistics of the selected failure modes as the reward function in the Multi-arm Bandit (MAB) problem \citep{zhang2020bandit}. However, different from \citep{zhang2020bandit}, the selection of arm is on the sensor level, where the SR statistics is defined on the failure-mode level. For high-dimensional data, specifying the joint distribution of high-dimensional data can be very challenging. Therefore, this paper will explore spatial structures for defining the failure mode distributions as shown in \ref{subsec: failure}. This paper also discussed that with the independence assumption of the distribution variable, the computational efficiency can be greatly improved.

The paper is organized as follows. In Section \ref{sec: review}, we will review the existing literature on change detection and isolation. We will also discuss works on process monitoring with resource constraints. We further introduce our proposed method and then discuss its property in Section \ref{sec: methods}. Then, we apply the proposed approach to both the simulated data and evaluate its performance and compare the existing methods in Section \ref{sec: simulation}. Furthermore, we apply the proposed method to two real cases in Section \ref{sec: casestudy}, respectively. Concluding remarks are given in Section \ref{sec: conslusion}.

\section{Literature Review} \label{sec: review}
In this section, we will provide a more detailed review of statistical process control or sequential change point detection methods. We will briefly classify the methods for the following four categories:  monitoring of independent HD streaming data, monitoring of functional data or profile monitoring, process monitoring with the resource constraint, change-point detection, and isolation.

In the first category, monitoring the HD streaming data has often been treated as monitoring the multiple independent univariate data streams. There are two distinct frameworks for monitoring independent data streams in recent years. First, the "global monitoring" framework focused on directly designing the global monitoring statistics for the process monitoring or change point detection for high-dimensional data \citep{xie2013sequential,wang2015large,cho2015multiple,chan2017optimal}. However, the global monitoring framework is typically computationally inefficient for high-dimensional data. Second, the "local monitoring and global decision" framework focus on monitoring each data stream independently using local monitoring statistics and then fusing these monitoring statistics together via the global statistics \citep{mei2010efficient,mei2011quickest}. The benefit is that these methods are typically computationally efficient and can be scalable to high-dimensional data. However, these methods are often limited to the independent data stream. Finally, this framework is targeted for the case of fully observed data, which may not be applicable under the resource constraint. 

In the second category, profile monitoring techniques have been proposed to tackle the complex spatial correlation structures.  Dimensionality reduction techniques, such as principal component analysis (PCA), are widely used. Various types of alternatives such as  multivariate functional PCA \citep{paynabar2015change}, multi-linear PCA \citep{grasso2014profile}, and tensor-based PCA \citep{yan2015image} are proposed.  On the other hand, non-parametric methods based on local kernel regression \citep{qiu2010nonparametric} and splines \citep{chang2010statistical} are developed.  To monitor the non-smooth waveform signals, a wavelet-based mixed effect model is proposed in \citep{paynabar2011characterization}. However, for both PCA-based methods and non-parametric methods, they typically assume that the change alternative is not known. To utilize the anomaly structures,  smooth sparse decomposition methods have been proposed and utilize two sets of basis functions, the background basis and anomaly basis, to represent the spatial structures of the background and anomaly, which have been applied to smooth profiles \citep{yan2017anomaly,yanRealTime2018} and waveform profiles \citep{yue2017wavelet}. However, all the profile monitoring techniques assume that the complete measurements are given and cannot be applied for HD data with partial observations. 

In the third category, many existing works focus on the change point detection with the sampling constraint. Here, we will briefly classify the existing monitoring methods with the sampling constraint into two categories, monitoring the i.i.d data stream and monitoring the correlated data stream. For monitoring the i.i.d data stream with the sampling constraint, \cite{liuAdaptive2015} proposed a top-R-based adaptive sampling strategy as a combination of random sampling in the in-control state and fixed sampling in the out-of-control state. Another work by \citep{zhang2020bandit} converts the problem into a MAB framework and adaptively selects the sensors with Thompson Sampling. Recent work by \citep{gopalan2021bandit} provides an information-theoretic lower bound for the detection delay. However, this method can be applied to multi-dimensional problems, but cannot be applied to the case with multiple complex failure-mode (i.e., after-change) distributions. 

However, due to the i.i.d assumption, these methods might not be suitable for data with complex distributions in reality. To deal with this problem, \cite{xianNonparametric2018} proposed an adaptive sampling strategy that can handle the correlated data generated from a multinomial distribution. For monitoring correlated data streams with the sampling constraint, these methods can be classified into monitoring data generated from the Bayesian Network and spatial profile. 
For example, \cite{liu2013adaptive} and \cite{liuObjectiveoriented2013} proposed a sensor allocation strategy according to a Bayesian Network to detect changes with multivariate $T^2$ control chart. Another work  discussed the problem when there is a spatial correlation among sensors and proposed a spatial-adaptive sampling strategy to focus on suspicious spatial clusters\citep{wangspatialadaptive2018}. However, these methods either consider the data stream with spatial correlation \citep{wangspatialadaptive2018,ren2020large,gomez2022adaptive} or modeled by the Bayesian Network structures \citep{liu2013adaptive,liuObjectiveoriented2013}, which fails to apply to the problem with general failure mode distributions as discussed in this paper. 

Finally, there is a large amount of work focused on the case when there are multiple failure modes, and it is necessary to identify the true failure while detecting the changes. The problem of sequential change detection with multiple failure modes is usually called change detection and isolation. The goal is to find the best decision procedure that can control the false alarm rate as well as the false isolation probability. The problem is of importance since it is common in different applications like fault diagnosis, process monitoring, and object identification \citep{nikiforov1993application, willsky1976survey, malladi1999generalized}. The major works in change detection and isolation can be categorized into Bayesian and non-Bayesian directions. \cite{nikiforov1995generalized} proposed a change detection/isolation framework as an extension of Lorden's results \citep{lorden1971procedures} which follows non-Bayesian schema. Another two works formulate the problem into a Bayesian version, which considers the change point as a random variable \citep{chen2020bayesian, malladi1999generalized}.

\section{Proposed Methodology} \label{sec: methods}

In this section, we will first describe the problem formulation of partially observed multi-mode change detection based on high-dimensional (HD) streaming data with sampling control in Section \ref{subsec: formulation}. We will review some relevant methodology in Section \ref{sec: reviewrelevent}. 
We will describe the proposed MTSSRP methodology in Section \ref{subsec: proposed}. 
We will prove important properties of the proposed algorithms about the average run length and failure mode isolation guarantee in Section \ref{subsec: property}. 
We will give the guidelines to select the tuning parameters of the proposed MTSSRP method in Section \ref{subsec: tuningparamers}. Finally, we will give a discussion and several guidelines on selecting the failure mode distributions in Section \ref{subsec: failure}.


\subsection{Problem Formulation and Background} \label{subsec: formulation}
Suppose we are monitoring data stream $X_{j,t}$ for $j=1,\cdots,p$ and $t=1, 2, \cdots,T$, where $p$ is the number of dimensionality in the system and $T$ is the monitored time length. We assume that the data streams follow joint distribution $f_{0}$ before change as ${\bf X}_{t} \sim f_0 $ for $t<\nu$. At some unknown change time $\nu\in\{1,2,\cdots,\}$, an undesirable event occurs and causes an abrupt change of the data stream into one or few failure modes. For example, the after-change distributions $f_k\in\mathcal{F}$ can be anyone from a family of distributions $\mathcal{F} = \{f_{1},\cdots,f_{K}\}$. In another word, after the change,  ${\bf X}_{t} \sim f_{k}, k=1,\cdots,K$ for $t>\nu$. 
In other words, we do not assume that we know which failure mode occurs in the system. Following the change detection and isolation framework, we do assume that a single failure mode $f_k$ may occur after the change. We will discuss the case with multiple failure mode. 
Here, $f_0, f_1, \cdots f_K$ are the joint distribution for all sensing variables. The sensing variables can also also be correlated. Finally, in practice, we can set the magnitude of the joint distribution as the interested magnitude of the change to be detected. 

Furthermore, we assume that given the resource constraint, it is not possible to observe all the data streams. For the partially observed data with sampling control, the set of the observed data is denoted by  $\mathbf{y}_t  = \{ X_{j,t}, j \in C_t \}$. Here, $C_t$ is the set of observed sensor indices at time $t$, which can be selected online. In  other words, we can define $a_{j,t}$ as the binary variable denoting whether the variable $j$ is observed at time $t$,  $C_t =\{j: a_{j,t} = 1\}$. Finally, the sampling constraint is represented by $\sum_{j=1}^p a_{j,t} = q$ at each time $t=1,2 \cdots $, which means at each time $t$, only $q$ sensing variables can be observed from all $p$ variables.  

The objective of this paper is to design an efficient adaptive sampling algorithm and the change point detection algorithm to automatically distribute sensing resources according to the knowledge of the system failure modes such that the change can be detected quickly as soon as it occurs and the corresponding failure mode can be identified accurately while maintaining the false alarm constraint. 

\section{Review of Relevant Methodology} \label{sec: reviewrelevent}

In Section \ref{sec: reviewrelevent}, we will review the formulations of the relevant methodology in detail.  
Section \ref{subsec: SRreview} reviews the Bayesian decision framework, namely the Shiyaev-Roberts (SR) procedure for sequential change-point detection methods. Section \ref{subsec: TSSRP} reviews the extension of the SR procedure for HD data monitoring. 

\subsection{Review of Shiyaev-Roberts statistics for uni-variate Sequential Change Point Detection} \label{subsec: SRreview}

We will first review the Bayesian decision approach for the sequential change-point detection by \cite{shiryaev1963optimum}. Consider a univariate streaming data ${\bf X}_1,{\bf X}_2\cdots,$ where the distribution of the data changed from $f_0$ to $f_1$ at some unknown time $\nu$. In another word, We assume that ${\bf X}_1,\ldots,{\bf X}_\nu \overset{i.i.d}{\sim}  f_0$ and ${\bf X}_{\nu+1},\ldots \overset{i.i.d}{\sim}  f_1$. The goal is to propose a statistical decision policy to determine the stopping time $T$, where $T=t$ implies that a change has happened at time $t$. 

Here, the goal is to find a decision policy such that the detection delay can be minimized under the false alarm constraint.
A Bayesian formulation on the statistical decision policy has been proposed, where the true change point $\nu$ is assumed to has geometric distribution as $P(\nu=t) = p(1-p)^{t-1}$. When $p \rightarrow 0$, \citep{shiryaev1963optimum} proposed a statistics $R_t=\sum_{j=1}^t \prod_{i=j}^t \frac{f_1({\bf X}_i)}{f_0({\bf X}_i)}$, which can be computed recursively as 
\begin{equation}
    R_t = (R_{t-1} + 1) \frac{f_1 ({\bf X}_t)}{f_0 ({\bf X}_t)}.
\end{equation}

For some pre-specified constant $A$, the decision to detect a change point in distribution can be made as
$
T_A=\inf_t \{R_t\ge A\}.
$
\cite{pollak1985optimal,pollak1987average} proved that this change point detection rules has an asymptotic minimax property, which is to minimize 
$
\sup_{1\le \nu<\infty}\mathbb E(T_A-\nu|T_A\ge \nu)
$
under the constraint that $\mathbb E(T_A|\nu=\infty)\ge B$ and $B\to\infty$. 

Despite the statistical efficiency, SR procedure is defined for the change point detection for univariate data, which cannot be used to monitor the high-dimensional data.

\subsection{Review of Thompson Sampling with the SR statistics} \label{subsec: TSSRP}


\citet{zhang2020bandit} combined the Thompson Sampling with the SR statistics to monitor high-dimensional with adaptive sampling control. Suppose for any time $t$, the data is $p$ dimensional. Before change time $\nu$, the $p$th dimension is $\overset{i.i.d}{\sim} f_{0}$, and after change time $\nu$, the $p$th dimension is  $\overset{i.i.d}{\sim} f_{1}$. \citet{zhang2020bandit} proposed a statistic for each dimension with
\begin{equation}
R'_{j,t}=\begin{cases}
(R'_{j,t-1}+1)\frac{f_{1}(X_{j,t})}{f_{0}(X_{j,t})} & j~{\rm observed}\\
R'_{j,t-1}+1 & j~{\rm unobserved}
\end{cases}
\label{eq: TSSRP}
\end{equation}
The decision of a change point of distribution is a threshold of top $r$ statistics, which is 
$$
T=\inf\{t\ge 1:\sum_{j=1}^r R'_{j,t}\ge A\}
$$
To determine which dimensions to observe, \citet{zhang2020bandit} implemented Thompson Sampling, where a random variable $R'_{j,0}$ is sampled from a prior distribution $G$ and added to $R'_{j,t}$ with some coefficients. The $q$ dimensions with the largest result are observed at time $t+1$. Even though this algorithm provided a theoretical understanding of the compensation terms in \citep{liuAdaptive2015} from the Bayesian perspective, it is still focused on the case where the failures occurred on individual sensor level, which can not be applied to HD data with complex failure mode distributions.

\subsection{Review of Change Point Detection and Isolation} \label{subsec: isolation}

We will briefly review the change-point detection and isolation framework by \cite{chen2020bayesian}.  Consider a streaming data ${\bf X}_1,{\bf X}_2\cdots,$ where the distribution of the data changed from $f_0$ to $f_k$ at some unknown time $\nu$. In another word, We assume that ${\bf X}_1,\ldots,{\bf X}_\nu \overset{i.i.d}{\sim}  f_0$ and ${\bf X}_{\nu+1},\ldots \overset{i.i.d}{\sim}  f_k$, where $f_k \in \mathcal{F} = \{ f_1, \cdots, f_K\}$ is a known pre-defined set of after-change distributions. 

The goal of a change-point detection and isolation algorithm is to compute a terminal pair $\delta = (T, \hat{k})$ online based on the observations ${\bf X}_1,{\bf X}_2,\cdots$, where $T$ is the alarm time at which $\hat{k}$-type change is identified.
The goal is to propose a statistical decision policy to determine the stopping time $T_A$, where $T_A=t$ implies that a change has happened at time $t$. Most of the change-point detection framework will use pre-specified constant $A$, the decision to detect a change point in distribution can be made as $T_A=\inf_t \{R_t\ge A\}.$ 

Similar to the change point detection problem, we would like to minimize the detection delay as 
$\sup_{1\le \nu<\infty}\mathbb E(T_A-\nu|T_A\ge \nu)$.

Besides, there are typically two types of constraints: 1) under the case that there is no change, the average detection time should be larger than a threshold $B$ as $\mathbb E(T_A|\nu=\infty)\ge B$. 2) If a true change mode $k$ happens, the false isolation rate, which is defined as: $\max\limits_{1\leq k\leq K}\mathbb P^k\{\hat{k}\neq k|T\geq\nu \}$ should also be small. However, these methods assume that the data is fully observable, which cannot be used to partially observed data.

\subsection{Proposed Algorithm}
\label{subsec: proposed}
\begin{figure}
    \centering
    \includegraphics[width=0.6\textwidth]{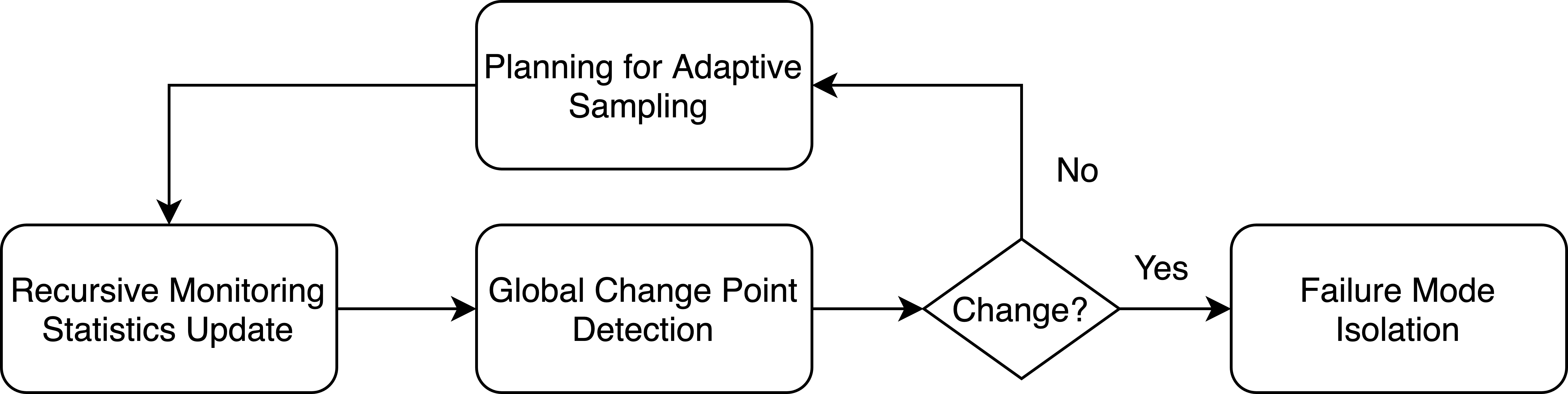}
    \caption{Procedure of the proposed method}\label{fig:framework}
\end{figure}

In Section \ref{subsec: proposed}, we will introduce the proposed methodology with the following major steps, monitoring statistics update, change point detection decision, failure mode isolation, and planning for adaptive sampling. The overall framework is shown in Fig.~\ref{fig:framework}, and the detailed steps are as follows:
\begin{enumerate}
\item \textbf{Monitoring statistics update}: We first construct the monitoring statistics of partially observed HD streaming data for each failure mode based on the SR procedure. The detailed step is discussed in Section \ref{subsubsec: recursiveMonit}. 
\item \textbf{Change point detection decision}: According to the updated monitoring statistics for each failure mode, a top-R statistic is used to conduct the global decision. We will then raise a global alarm if the process has gone out of control and decide which failure mode has occurred. The detailed step is discussed in Section \ref{subsubsec: detection}.
\item \textbf{Planning for adaptive sampling}: If the change is not detected, we will update the sampling layout dynamically according to the historical observations. To achieve this, we propose to borrow the Thompson sampling idea to decide the next sampling layout, where the data is randomly sampled from the identified failure mode distribution. The detailed step is discussed in Section \ref{subsubsec: planning}. Furthermore, the optimization algorithm to solve this planning and optimal sampling decision is discussed in Section \ref{subsubsec: optimization}. The selected sampling patterns will be used to update the monitoring statistics recursively. 
\item \textbf{Failure mode isolation}: Finally, if the change is detected, we will isolate and identify the true failure mode in the system. 
\end{enumerate}

\subsubsection{Recursive Monitoring Statistics Update} \label{subsubsec: recursiveMonit}
In this subsection, we will discuss the proposed method of constructing the Shiyaev-Roberts (SR) statistics for each failure mode $k\in\{1,\cdots,K\}$ with missing observations. Here, we denote the local SR statistics at time $t$ as $R_{k,t}$. Here, $\mathbf{y}_{t}$ is the set of observed data streams. We will follow the same rule of updating the local statistics $r_{k,t}$. 

\[
R_{k,t}=(R_{k,t-1}+1)\frac{\tilde{f}_{C_t, k}(\mathbf{y}_{t})}{\tilde{f}_{C_t,0}(\mathbf{y}_{t})}. 
\]
Here, $\tilde{f}_{C_t, k}$ is the joint distribution of the observed data  $\mathbf{y}_{t}$ at set $C_t$. For computational efficiency and stability, it is recommended to use the $r_{k,t}=\log R_{k,t}$, updated as
\[
r_{k,t}=\log(\exp(r_{k,t-1})+1)+\log \frac{\tilde{f}_{C_t, k}(\mathbf{y}_{t})}{\tilde{f}_{C_t,0}(\mathbf{y}_{t})}.
\]
We set $R_{k,0}= 0 $ initially and update the statistics accordingly.  

\subsubsection{Detection Decision and Failure Mode Isolation} \label{subsubsec: detection}
We will combine the local statistics for each failure mode to construct a global stopping time. Here, if we know that the system only contains one failure mode, we propose to use the largest of the monitoring statistics $r_{k,t}$ to trigger the alarm. For example,  if $r_{k,t}$ is larger than a control limit $A$, to raise the alarm. 

\begin{equation}
    T = \inf\{t\geq 1: \max\limits_{k} r_{k,t}\geq A\}
\end{equation}
However, if we know that there are multiple failure modes in the system, the summation of the top $r_{k,t}$ statistics can be used to trigger the alarm. 
\begin{equation}
    T = \inf\{t\geq 1: \sum_{k=1}^{K_s} r_{(k),t}\geq A\}
\end{equation}

Finally, to isolate the most probable failure mode when the change is detected, we propose to use the monitoring statistics with the largest index  $\hat{k}$, computed as
\begin{equation}
\hat{k} =  \arg\max\limits_{k} r_{k,T}.
\end{equation}

\subsubsection{Planning for Adaptive Sampling}
\label{subsubsec: planning}
In this section, we present an efficient method to plan and select the best sampling pattern to observe at the next time point. Suppose that we have observed $\mathbf{y}_1,\cdots,\mathbf{y}_{t-1}$ now and the goal is to determine $\{a_{j,t} \}$ at next time $t$, which is a binary variable denoting whether  variable $j$ is observed or not at time $t$. 
Inspired by the MAB, we propose to maximize the reward function, defined by the monitoring statistics of the top few selected failure modes. More specifically, we propose to use the summation of the SR statistics of the top-$K_s$  failure modes as the reward function, where $K_s$ is a pre-defined parameter to balance the exploration and exploitation.  
In  other words, we can define the reward function as $S_{t} = \sum_{k=1}^{K_s} r_{(k),t} $, where $r_{(k),t} $ is the rank of the statistics such as $r_{(1),t}\geq\cdots\geq r_{(K_s),t}$. 
One specific challenge is that to compute the reward function $S_{t}$ for planning, we need to compute $r_{(k),t}$, which requires $\mathbf{x}_{t}$ to be fully observed. However, given that we are still at time $t-1$ yet and data $x_t$ has not been observed yet, it is impossible to compute and optimize $S_t$ for the planning problem. 

To solve this planning problem, we propose to optimize a sampled version of the monitoring statistics $S_t$, defined as $\tilde{S}_t= \sum_{k=1}^{K_s} a_{k,t}r_{(k),t}$. Borrowing from the Thompson sampling algorithm, we would like to use the sampled version of $\mathbf{x}_t$, denoted as $\mathbf{\tilde{x}}^k_{t}$ as
\begin{equation}
\tilde{r}_{k,t}=\log(\exp{(r_{k,t-1})}+1)+ \log \frac{f_{k}(\tilde{\mathbf{x}}^k_{t} )}{f_{0}(\tilde{\mathbf{x}}^k_{t} )}, 
\end{equation}
where $\tilde{\mathbf{x}}^k_{t} \sim f_k$ is sampled from the $k^{th}$ failure mode. Finally, one can optimize $a_{j,t}$ by maximizing the sampled version of $S_{t}$, denoted as $\tilde{S}_{t}$ as 
 
\begin{align}
 & \max_{a_{j,t}}\tilde{S}_{t} \quad \text{subject to} & \sum_{j}a_{j,t}=q,a_{j,t}=\{0,1\}  \label{eq: OptA}  
\end{align}
Finally, we will discuss how to solve the optimization (\ref{eq: OptA}) in Section \ref{subsubsec: optimization}. 

\subsubsection{Optimization for Planning}
\label{subsubsec: optimization}

In general, optimizing (\ref{eq: OptA}) is often challenging. If the problem dimension $p$ and the number of selected sensing variables $q$ are small, we can enumerate all the $ \tbinom{p}{q}$ possible combinations of the sampling layouts. However, given the time complexity is $O\left( \tbinom{p}{q} \right)$, enumeration of all possible combinations is not feasible for large $p,q$.

Here, we will first present the closed-form solution for a special case of the proposed algorithm, where the joint failure mode distribution $f_k(\mathbf x_t) = \prod_j f_{j,k}  (x_{t,j})$ can be approximated by independent but not necessarily identical distributions in each dimension $j$. Notice that if the possible failure modes in each dimension $j$ is finite, the total number of possible failure modes for all dimensions is finite. Here, we would like to derive the analytical solution to optimize (\ref{eq: OptA}) under this setting in Proposition \ref{thm: sort}.

\begin{prop} \label{thm: sort} If $f_k(\mathbf x_t) = \prod_j f_{j,k}  (x_{t,j})$ for $k=0,\cdots, K$, the set $C_t$ in (\ref{eq: OptA})  can be solved by selecting the indices of the largest $q$ of $s_{t,j}$, denoted as $s_{t,(1)}, s_{t,(2)}, \cdots, s_{t,(q)}$. 
Here $s_{t,j}$ is defined as
\begin{equation}
s_{t,j}= \sum_{k=1}^{K_{s}} \log \frac{f_{j, (k)}(\tilde{x}^k_{j, t})}{f_{j,0}(\tilde{x}^k_{j, t})}  \label{eq: scorecompute}
\end{equation}
  $s_{t,(j)}$ is the order statistics, defined as $s_{t,(1)} \geq s_{t,(2)} \geq \cdots \geq s_{t,(q)} \geq s_{t,(q+1)} \geq \cdots \geq s_{t,(p)}$. 
 \end{prop}
 We would like to mention that the computation of (\ref{eq: scorecompute}) in Proposition \ref{thm: sort} is actually very efficient. To compute each $s_{t,j}$, it requires the summation of $K_s$ terms, which is of $O(K_s)$ complexity. To compute all $p$ sensing variables at each time $t$, it requires only $O(p K_s)$ complexity at each time to decide the best sampling layout. Here, the limitation is that $f_{k,j}$ is assumed to be independent over different data dimension $j$. However, we find that even the distribution of each failure mode is not independent, this approximation can still achieve a pretty reasonable solution. 

In this paper, we will only focus on the monitoring of continuous variables and assume that the data follows a normal distribution $f_k \sim N(\boldsymbol{\mu}_k, \Sigma_k)$. However, as derived in the proposed framework, this method can be generalized to other distributions quite easily. 
Finally, as mentioned in Proposition \ref{thm: sort}, if we will further assume that $\Sigma_k$ is diagonal as $\Sigma_k = diag(\sigma_{k,1}^2,\cdots \sigma_{k,p}^2)$, we can derive a simpler formula for $s_{t,j}$ in Proposition \ref{GaussianCase}. 

\begin{prop} \label{GaussianCase}
Given that  $f_{k,j} = N(\mu_{k,j}, \Sigma_k )$, where  $\Sigma_k = diag(\sigma_{k,1}^2,\cdots \sigma_{k,p}^2)$. We can derive $s_{t,j}  = \sum_{k=1}^{K_{s}} (\frac{1}{\sigma_{k,j}^{2}}(\tilde{x}^k_{j, t}-\mu_{k,j})^{2}-\frac{1}{\sigma_{k,0}^{2}}(\tilde{x}^k_{j, t}-\mu_{k,j})^{2})$, $\tilde{x}_t \sim f_k$, 
\end{prop}
We would like to emphasize that the independence assumption of each failure mode distribution is actually not required for the proposed algorithm. It is only useful to derive the closed-form solution in solving  (\ref{eq: OptA}). If the spatial dimension is not independent for different failure modes, the proposed planning procedure can still be optimized without the assumptions by approximating the optimal solution.  Here, we propose a greedy algorithm to detect the $a_{j,t}$ sequentially. The detailed step is given as follows. First, we can select the first sensing variable $j_1$ to optimize the sampled version $\tilde{S}_t$ by $j_1 = \arg \max \tilde{S}_t, \sum_j a_{j,t} = 1, a_{j,t}=\{0,1\}.$ After $j_1$ is decided, we would like to choose the second sensing variable $j_2$ by 
$j_2 = arg \max \tilde{S}_t, \sum a_{j,t}=2, a_{j_1,t}=1, a_{j,t}=\{0,1\}$. We will continue the procedure until $j_q$ is selected. In conclusion, the set of the observed sensing index is given as $C_t = \{j_1, \cdots, j_q \}$. Given that we only need to enumerate all $p$ dimensions in each of the $q$ iterations, the time complexity can be reduced to $O(pq)$. Despite the efficiency, the greedy forward selection strategy usually does not produce a global optimal solution.

Here, we would like to highlight the major difference of the proposed method compared to the existing literature of monitoring of the i.i.d data stream such as \citep{zhang2020bandit}: 1)  the number of failure modes $K$ does not need to be the same as the  dimensionality $p$ of the data stream; 2) The normal data distribution $f_0$ does not need to be i.i.d according to each dimension as $x_{j}\overset{i.i.d}{\sim}f(x), \text{for all } j$; 3) For each failure mode, it can include overlapping sets of sensing variables.

Finally, we would like to point out a special version of the proposed method and how it links to  \citep{zhang2020bandit}, if we are interested in monitoring the i.i.d data stream with the focus of detecting the change of each individual sensing variable.

\begin{prop} \label{iid}
For before change  $H_0: x_{j}\overset{i.i.d}{\sim}f(x), \text{for all } j$. After change, for $j^{th}$ failure mode, where $j\in\{1,\cdots,p\}$, only 1 distribution changed the distribution to $g(x)$ as $x_{j}{\sim} g(x)$, where the rest $x_{j'}  \overset{i.i.d}{\sim} f(x), j'\neq j$ still follows the pre-change distribution. The proposed algorithm will result in the sampled updating rule as in \citep{zhang2020bandit}:
\begin{equation}
R_{j,t}=\begin{cases}
\frac{g(\tilde{x}_{j,t})}{f(\tilde{x}_{j,t})}(R_{j,t-1}+1) & a_{j,t}=1\\
R_{j,t-1}+1 & a_{j,t}=0
\end{cases}
\end{equation}
\end{prop}
Proposition \ref{iid} shows a special case for the proposed algorithm, which assumes the change only affected a few data streams and the algorithms try to identify the change with the resources constraint. Under the current setting, the proposed algorithm will become another sampled version of the TSSRP algorithm. Many previous works, including \citep{liuAdaptive2015,zhang2020bandit} have studied this setting. However, the proposed algorithm can be generalized into any other joint distributions of different failure modes. 
\subsection{Properties of the Proposed Algorithm} \label{subsec: property}

Here, we will prove two important properties of the algorithms about the bound of the average run length and the failure mode isolation in Theorem \ref{Run} and Theorem \ref{Limit}, respectively. 

\begin{theorem}[Average Run Length]\label{Run}
Let $T=\inf\{t\geq 1: r_{(1),t} \geq A_2 \}$. Then we have that under the null hypothesis where no changes occur, $\mathbb{E} T\ge A_1/K$, $\mathbb{E} T=O(A_1)$, where $A_1=e^{A_2}$. 
\end{theorem}
 Theorem \ref{Run} provides a lower and upper bound for the Average Run Length if no changes occur. Theorem \ref{Run} provides us the guidance to select conservative upper and lower bounds of the control limit $A$. Specifically, $K * ARL$ can serve as the upper bound in the bisection search to speed up the threshold choosing procedure.

\begin{theorem}[Failure Mode Isolation]\label{Limit}
Assume ${\bf X}_1\ldots {\bf X}_\nu\sim f_0$, ${\bf X}_{\nu+1}\ldots\sim f_k$. All distributions of failure modes are continuous, and the KL divergence of the distributions of failure mode $l$ and the true failure mode $k$ follows $0<KL(f_k \|f_l)<\infty$, and $\mathrm{Var}_{x\sim f_k} [\log \frac{f_k}{f_l}]<\infty$, $\text{for all } l\neq k$. The probability that $P(r_{k,t} > r_{l,t} ) \rightarrow 1$ as $t\to\infty$.
\end{theorem}

Theorem \ref{Limit} provides the behavior of the largest SRP statistics when time goes to infinity under the alternative hypothesis (where the failure mode $k$ occurs). It shows that the adaptive sampling algorithm will always be able to isolate the true failure mode $k$ if $t\rightarrow \infty$.

In some cases, there might be multiple failure modes happening at different time points after the change point $\nu$, i.e. for some $t_1>\nu$, ${\bf X}_{t_1}\sim f_k$; for some $t_2>\nu$, ${\bf X}_{t_2}\sim f_l$. Corollary \ref{limit-cor} is proved. 

\begin{corollary}\label{limit-cor}
Assume ${\bf X}_1\ldots {\bf X}_\nu\sim f_0$. Let $\mathcal K=\{k: {\bf X}_t\sim f_k\textit{ for some }t>\nu\}$ be the set of true failure modes. If we further assume that the support of different failure modes are non-overlapping, then for any $k\in \mathcal K$ and $l\notin \mathcal K$, we have $\lim_{t\to\infty}P(r_{k,t}>r_{l,t})=1$.
\end{corollary}
Corollary \ref{limit-cor} assumes that different failure modes are not overlapping with each other, we can prove in Corollary \ref{limit-cor} that SRP statistics of the true failure modes will be larger compared to those of the other potential failure modes. We would like to point out that Corollary \ref{limit-cor} is not always true for failure mode distributions that are potentially overlapped with each other. For example, if half of the data after the change follows $f_1$ and the other half after the change follows $f_2$, it might be possible that the true failure mode identified would be $f_3 = \frac{1}{2} (f_1 + f_2)$.

\subsection{Choice of Parameters} \label{subsec: tuningparamers}
Here, we will present practical guidelines for tuning parameter selection. Given that the number of sensing variables $q$ typically depends on the available sensing resources in the particular applications, we only need to select the following parameters:  the number of top-R selected failure modes $K_s$, the control limit threshold $A$, and the failure mode distribution $f_k$ and $f_0$.

\textbf{Choice of the number of observed failure modes $K_s$:} First, the number of selected failure modes for the monitoring statistics should be smaller than the total number of potential failure modes. Ideally, $K_s$ should be chosen as large as the total number of true failure modes in the system. In practice, we found that increasing $K_s$ to be more than the true number of failure modes in the system would lead the algorithm to explore more potential failure modes or increase the exploration power. However, if $K_s$ is too large, the algorithm is not able to focus on the actual failure modes, which decreases the exploitation power. 

\textbf{Choice of threshold $A$:} The choice of control limit $A$ can be determined by the In-control ARL (or $ARL_0$). If $A$ is large, $ARL_0$ will also increase. In practice, we can set an upper bound of the $A$  by utilizing the Theorem \ref{Run} and then use the binary search algorithm to find the best $A$ for a fixed $ARL_0$. 

\subsection{Selection of failure mode distribution $f_k$} \label{subsec: failure} 

Finally, the complex joint distributions of different failure modes also bring significant computational complexity, which will be addressed in this paper. 

Selecting the failure mode distribution is very important to achieve better change detection and isolation performance. However, it is very challenging to provide accurate failure modes definition for high-dimension data without any domain knowledge. In this work, we mainly focus on detecting mean-shift in high dimension data. We further discuss how to define the failure modes based on our knowledge of the high-dimension data. Overall, there are two strategies for choosing the most appropriate failure mode distribution $f_k$. 1) If there is prior knowledge about the failure mode distributions, we can set the distribution according to the prior knowledge. For example, if we know that the hot spots are clustered, each failure mode distribution can be assumed as the mean-shift of the IC distribution $f_0$ with an individual B-spline or Gaussian kernel basis. If we know that the post-change distribution is sparse, a simple way is to set the failure mode distribution as the mean shift of each individual sensor. 2) If we do not know the failure mode distributions, we can collect some samples for each failure mode and use these samples to estimate the failure mode distribution. 

\section{Simulation Study} \label{sec: simulation}
Here, we will evaluate the proposed method in a simulation study. 
We will start with the simulation setup for single failure mode in Section \ref{subsec: simsetup} with two different scenarios: the non-overlapping case and the overlapping case. Then, we will evaluate the proposed algorithm in these two scenarios. To test the robustness of the proposed algorithm in the case when multiple failure modes coexist, we also perform the sensitivity analysis to evaluate the performance in \ref{subsec: simu_muti}

\subsection{Simulation Study for a Single Failure Mode \label{subsec: simsetup}}
Here, we will discuss the two scenarios for the simulation setup. We are trying to distinguish multiple failure modes by whether these failure modes have overlapping support. For example, for the first "non-overlapping" case, we assume that different failure modes $f_i$ and $f_j$ have non-overlapping support.
\subsubsection{Scenario 1: the non-overlapping case}
We will first discuss the non-overlapping case of the proposed method. In the simulation, we let the data dimension $p=1000$, the number of failure mode $K = 50$, each time the algorithm will select $q=10$ sensors at each time. Here, we assume the normal data or in control (IC) data follows  $f_0 = N(0,I)$. After the change, the out-of-control (OC) data have $K$ failure modes, where $f_k = N(\boldsymbol{\mu}_k, I)$ and $\boldsymbol{\mu}_k = \sum_{j=sk}^{(s+1)k} e_j$.  $e_j=(0,\cdots,1,\cdots,0)$, and only $j^th$ element is 1. Here, three failure modes have been selected after the change. 
In other words, if we organize the 1000 data streams into a $50\times20$ image, each failure pattern would be each row of pixels which can be visualized in Fig.~\ref{fig:sim_iid_potential}.  Therefore, different failure modes are not overlapped, given they contain different sensing variables.  

\begin{figure}
    \centering
        \subfigure[All 50 Non-overlapping potential failure modes]{\includegraphics[width=0.45\textwidth,height=0.45\textwidth]{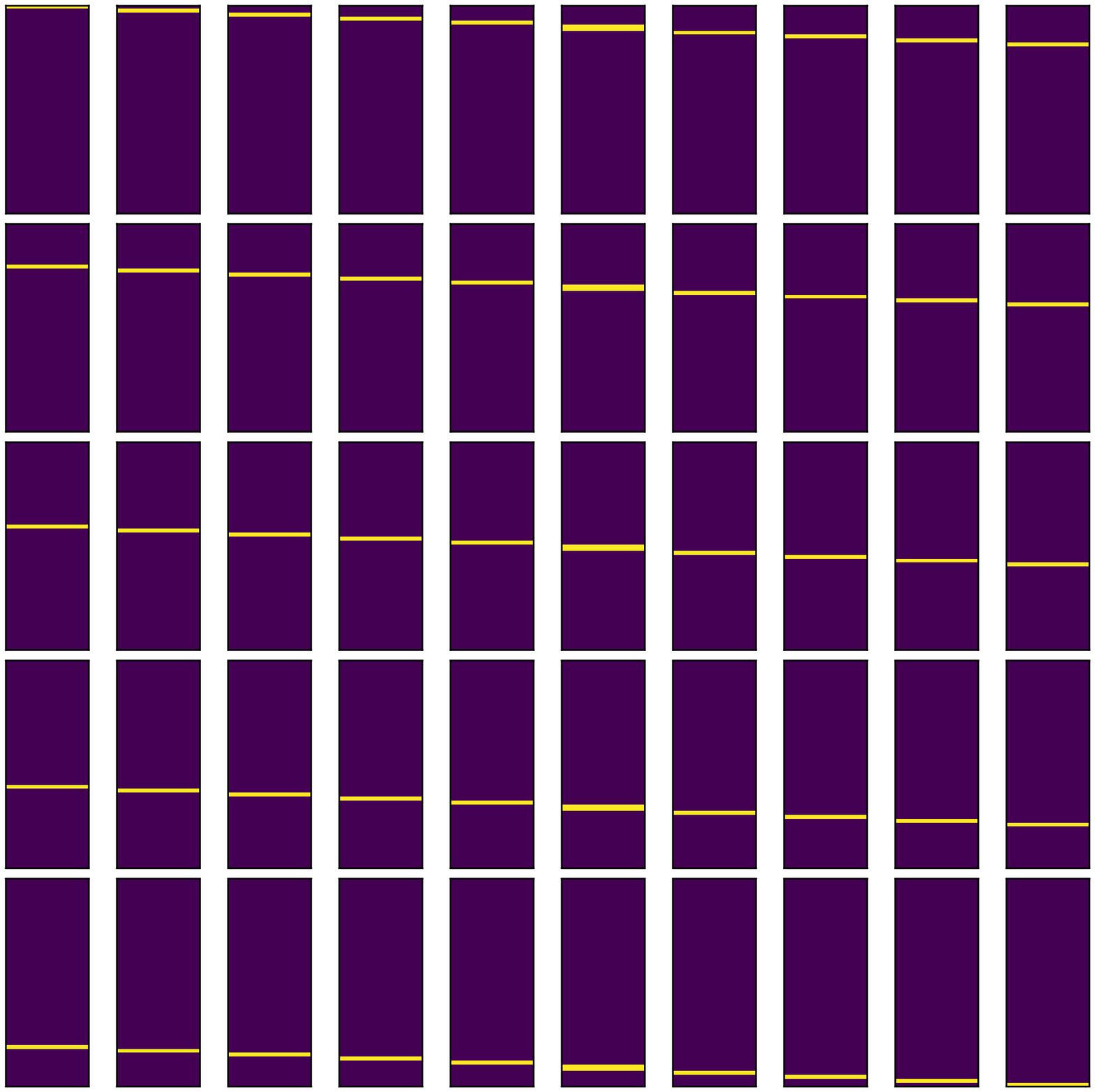} \label{fig:sim_iid_potential}} \subfigure[All 49 overlapping potential failure modes]{\includegraphics[width=0.45\textwidth,height=0.45\textwidth]{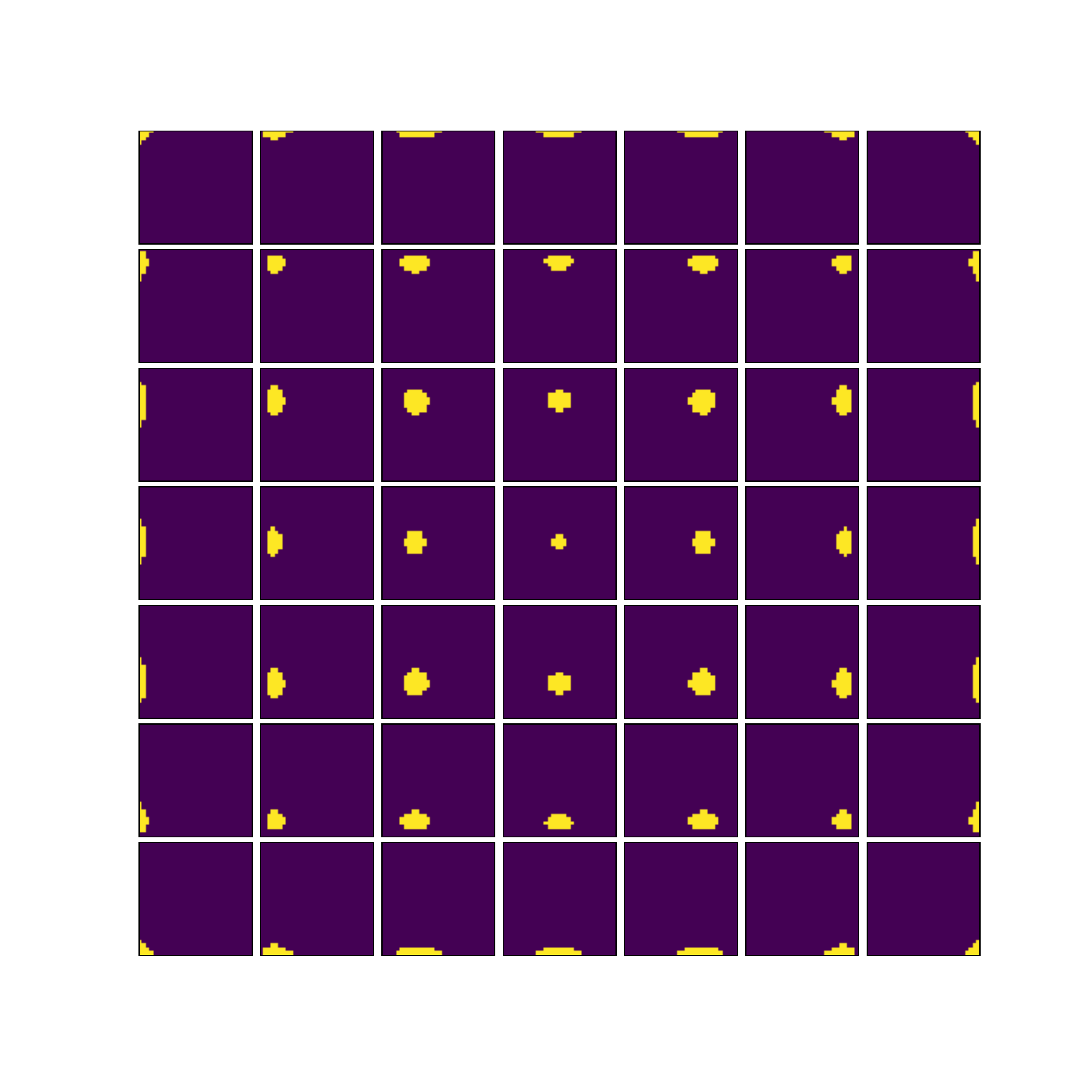}
    \label{fig:sim_spline_potential}}
    \caption{Failure Modes for Overlapping and Non-overlapping Cases}\label{fig:sim_anomaly}
\end{figure}

Finally, we assume that we are only accessible to $10$ out of $p=1000$ data streams to observe. At each time step, we adaptive select data streams to monitor the whole process.  We want to detect the correct failure mode as soon as possible. 

\subsubsection{Scenario 2: the overlapping case}
We will discuss the second scenario, where the failure patterns are generated as small spatial clusters. In this case, we set the data as 2-D images with size $30\times30$ with total dimension $p=900$. Here, the failure modes are generated using B-spline basis with $7$ knots in both $x$ and $y$ directions. As shown in Fig.~\ref{fig:sim_spline_potential}, we end up with $7^2=49$ potential failure modes. After the change happens, we randomly select a failure mode as the true failure mode. In this situation, some failure modes might overlap with each other, which will be more challenging for the algorithm to isolate the real changes. Finally, during the monitoring process, we can select 10 out of $p=900$ data streams adaptively to observe online. 

\subsubsection{Simulation Result \label{subsec: simresult}}
Here, we will compare the proposed MTSSRP with the following benchmark methods: 1) TSSRP method \citep{zhang2020bandit}, which is introduced in detail in Appendix. 2) TRAS method \cite{liuAdaptive2015}, where the local CUSUM statistics are used for each individual data stream and later fused together via the Top-R rule. To show the upper-bound and lower-bound performance, we will also add three simple alternatives: 1) Random, where we randomly select $q$ sensors at each time step with the top-r statistics by monitoring each sensor individually. 2) Oracle, where we have not only access to all the data streams but also the failure mode distribution information using the monitoring statistics as MTSSRP. 3) MRandom, where we apply the same monitoring statistics as MTSSRP, which considers the failure mode distribution information in the monitoring statistics, but we randomly select the sensors at each time step. We evaluate the proposed method with two metrics which are detection delay and failure isolation accuracy. 

\begin{figure}
\vspace*{-1.5cm}
    \centering
    \subfigure[$ARL_1$ in Non-overlapping Cases]{
\includegraphics[width=0.45\textwidth,height=0.36\textwidth]{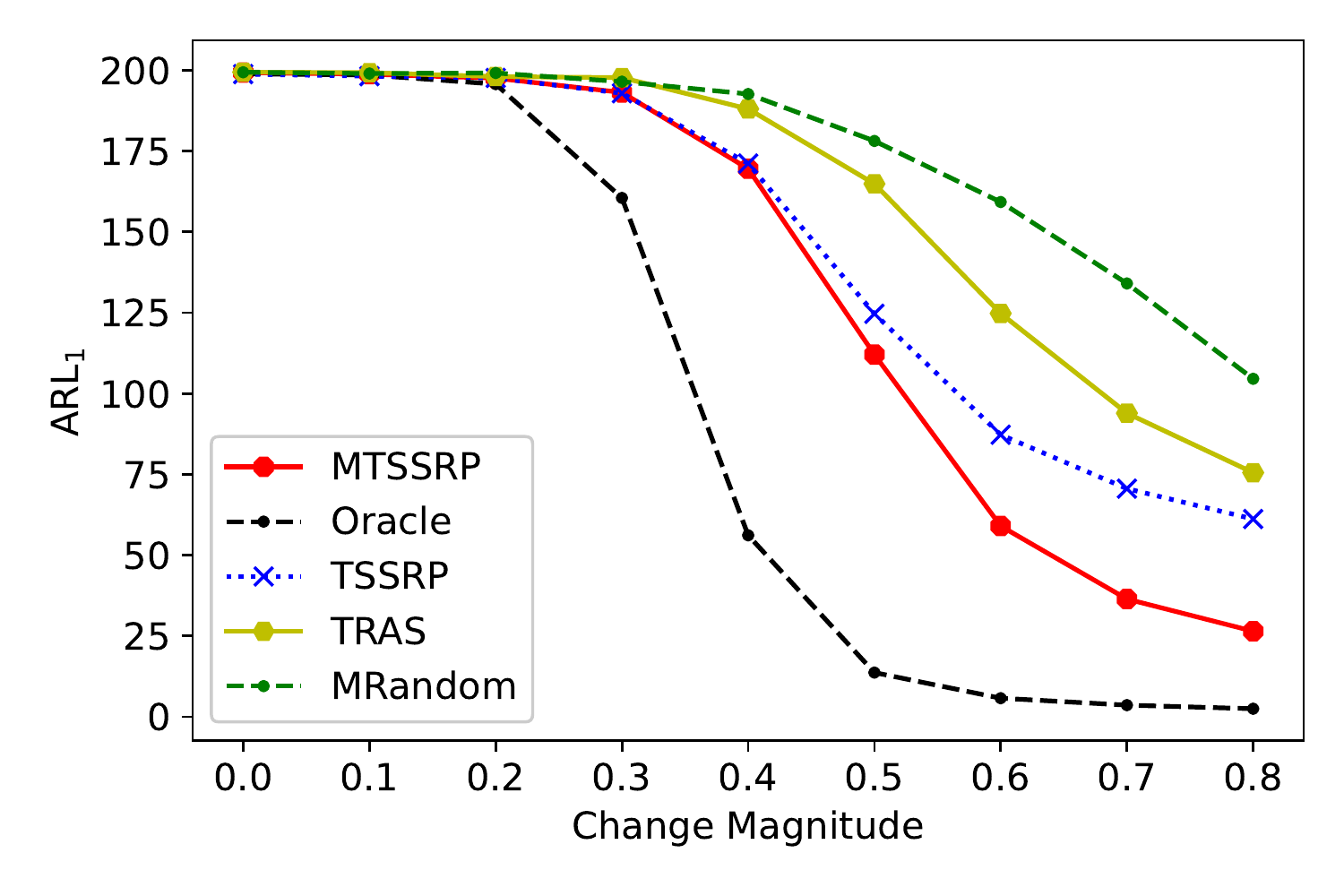}
    \label{fig:ARL1}}
    \subfigure[$ARL_1$ in Overlapping Cases]{
  \includegraphics[width=0.45\textwidth,height=0.36\textwidth]{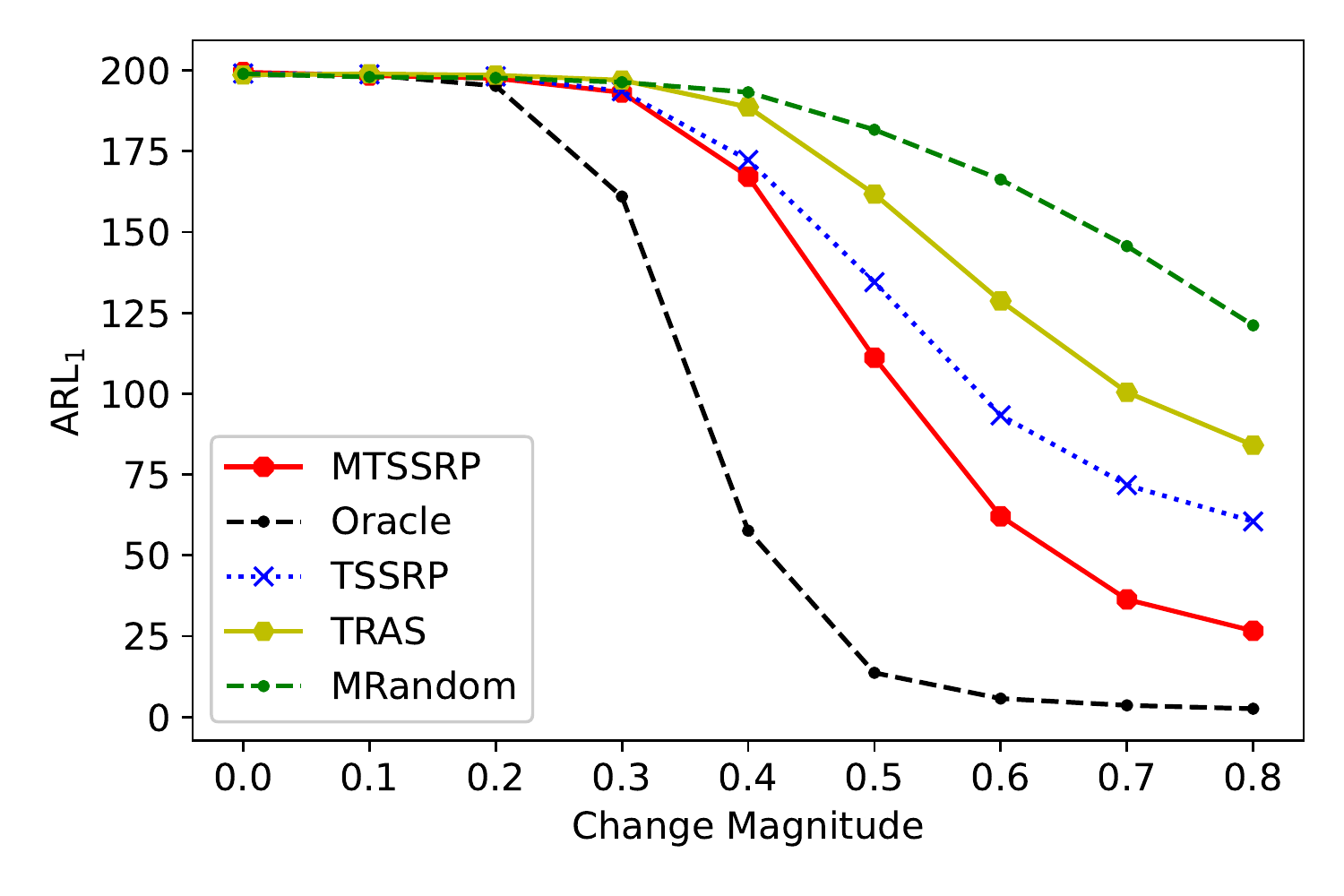}
    \label{fig:ARL2}}
    
    \subfigure[Accuracy in Non-overlapping Cases]{
\includegraphics[width=0.45\textwidth,height=0.36\textwidth]{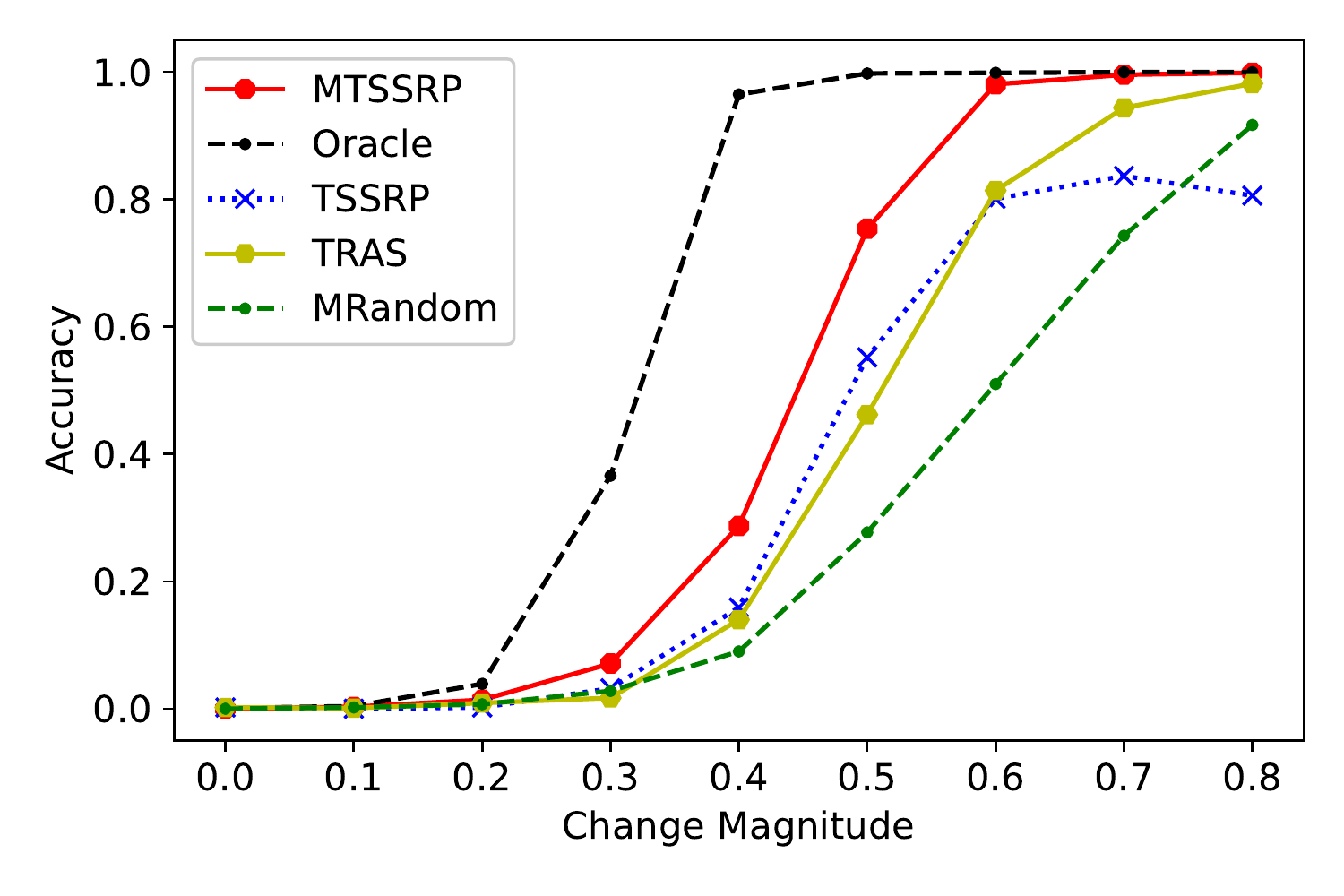}
    \label{fig:accuracy1}}
    \subfigure[Accuracy in Overlapping Cases]{
  \includegraphics[width=0.45\textwidth,height=0.36\textwidth]{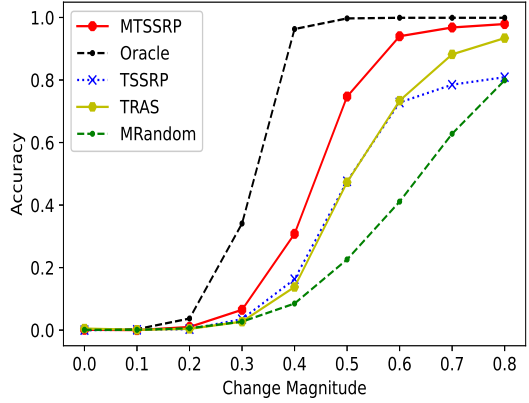}
    \label{fig:accuracy2}}
    \caption{Out-of-control Average Run Length ($ARL_1$) and Failure isolation accuracy for Different Change Magnitude $\delta$} \label{fig:ARL_single}
\end{figure}

First, we would like to compare the detection delay of the proposed method and all the benchmark methods. Here, we set the in-control average run length (i.e., denoted as $ARL_0$) for all methods as $200$ and compare their out-of-control ARL or average detection delay  (i.e., denoted as $ARL_1$) with 1000 replications. Here, we will compute the $ARL_1$ for different change magnitudes ($\delta$ = $0.5$, $0.8$) in Table \ref{table: ARLsim_single}. We also compared the $ARL_1$ and Isolation Accuracy from different magnitude $\delta$ ($0.1$ to $0.8$) in Fig.~\ref{fig:ARL_single}.
From the results, we can see that the proposed MTSSRP has better $ARL_1$ compared to other benchmark methods. MTSSRP performs much better than MRandom, which validates the efficiency of the proposed sampling strategy. The advantage of MTSSRP over TSSRP shows that considering the failure mode information can greatly improve the performance. We further compare the isolation accuracy as shown in Fig.~\ref{fig:accuracy1} and Fig.~\ref{fig:accuracy2}. It can be seen that the proposed MTSSRP achieves better performance than others. It can also reach pretty high accuracy when $\delta$ is greater than 0.6. 


\begin{table}
\centering \caption{Average Run Length and Failure Mode Isolation Accuracy for single failure}
\begin{adjustbox}{width=0.95\textwidth} %

\begin{tabular}{|cc|cccc|cccc|}
\hline
\multicolumn{2}{|c|}{\emph{Case}} & \multicolumn{4}{c|}{nonoverlap} & \multicolumn{4}{c|}{overlap} \\ \hline
\multicolumn{2}{|c|}{\emph{Change Magnitude}} & \multicolumn{2}{c|}{$\delta=0.5$} & \multicolumn{2}{c|}{$\delta=0.8$} & \multicolumn{2}{c|}{$\delta=0.5$} & \multicolumn{2}{c|}{$\delta=0.8$} \\ \hline
\multicolumn{2}{|c|}{\emph{Metrics}} & \multicolumn{1}{c|}{$ARL_1$} & \multicolumn{1}{c|}{Accuracy} & \multicolumn{1}{c|}{$ARL_1$} & Accuracy & \multicolumn{1}{c|}{$ARL_1$} & \multicolumn{1}{c|}{Accuracy} & \multicolumn{1}{c|}{$ARL_1$} & Accuracy \\ \hline
\multicolumn{2}{|c|}{\emph{Oracle}} & \multicolumn{1}{c|}{13.71(11.56)} & \multicolumn{1}{c|}{0.99(0.04)} & \multicolumn{1}{c|}{2.52(1.08)} & 1.0(0.0) & \multicolumn{1}{c|}{13.68(12.05)} & \multicolumn{1}{c|}{1.0(0.05)} & \multicolumn{1}{c|}{2.57(1.43)} & 1.0(0.03) \\ \hline
\multicolumn{1}{|c|}{\multirow{5}{2cm}{\emph{Competing Methods}}} & MTSSRP & \multicolumn{1}{c|}{\textbf{112.05(66.96)}} & \multicolumn{1}{c|}{\textbf{0.75(0.43)}} & \multicolumn{1}{c|}{\textbf{26.49(17.09)}} & \textbf{0.99(0.03)} & \multicolumn{1}{c|}{\textbf{111.16(66.01)}} & \multicolumn{1}{c|}{\textbf{0.75(0.43)}} & \multicolumn{1}{c|}{\textbf{26.66(17.71)}} & \textbf{0.98(0.14)} \\ \cline{2-10} 
\multicolumn{1}{|c|}{} & TSSRP & \multicolumn{1}{c|}{124.66(58.2)} & \multicolumn{1}{c|}{0.55(0.50)} & \multicolumn{1}{c|}{61.19(30.67)} & 0.81(0.39) & \multicolumn{1}{c|}{134.51(58.38)} & \multicolumn{1}{c|}{0.48(0.5)} & \multicolumn{1}{c|}{60.5(35.92)} & 0.81(0.39) \\ \cline{2-10} 
\multicolumn{1}{|c|}{} & TRAS & \multicolumn{1}{c|}{164.84(49.07)} & \multicolumn{1}{c|}{0.46(0.50)} & \multicolumn{1}{c|}{75.53(42.51)} & 0.98(0.13) & \multicolumn{1}{c|}{161.75(51.48)} & \multicolumn{1}{c|}{0.47(0.5)} & \multicolumn{1}{c|}{84.08(50.73)} & 0.93(0.25) \\ \cline{2-10} 
\multicolumn{1}{|c|}{} & MRandom & \multicolumn{1}{c|}{178.1(41.49)} & \multicolumn{1}{c|}{0.28(0.45)} & \multicolumn{1}{c|}{104.55(48.71)} & 0.92(0.28) & \multicolumn{1}{c|}{181.67(39.91)} & \multicolumn{1}{c|}{0.23(0.42)} & \multicolumn{1}{c|}{121.16(54.09)} & 0.8(0.4) \\ \cline{2-10} 
\multicolumn{1}{|c|}{} & Random & \multicolumn{1}{c|}{199.97(1.07)} & \multicolumn{1}{c|}{-} & \multicolumn{1}{c|}{199.66(5.08)} & - & \multicolumn{1}{c|}{200.0(0.0)} & \multicolumn{1}{c|}{-} & \multicolumn{1}{c|}{199.95(1.61)} & - \\ \hline
\end{tabular}\label{table: ARLsim_single}

\end{adjustbox} 
\end{table}

\begin{figure}
    \centering
    \subfigure[Potential Failure Mode]{\includegraphics[width=0.4\textwidth,height=0.3\textwidth]{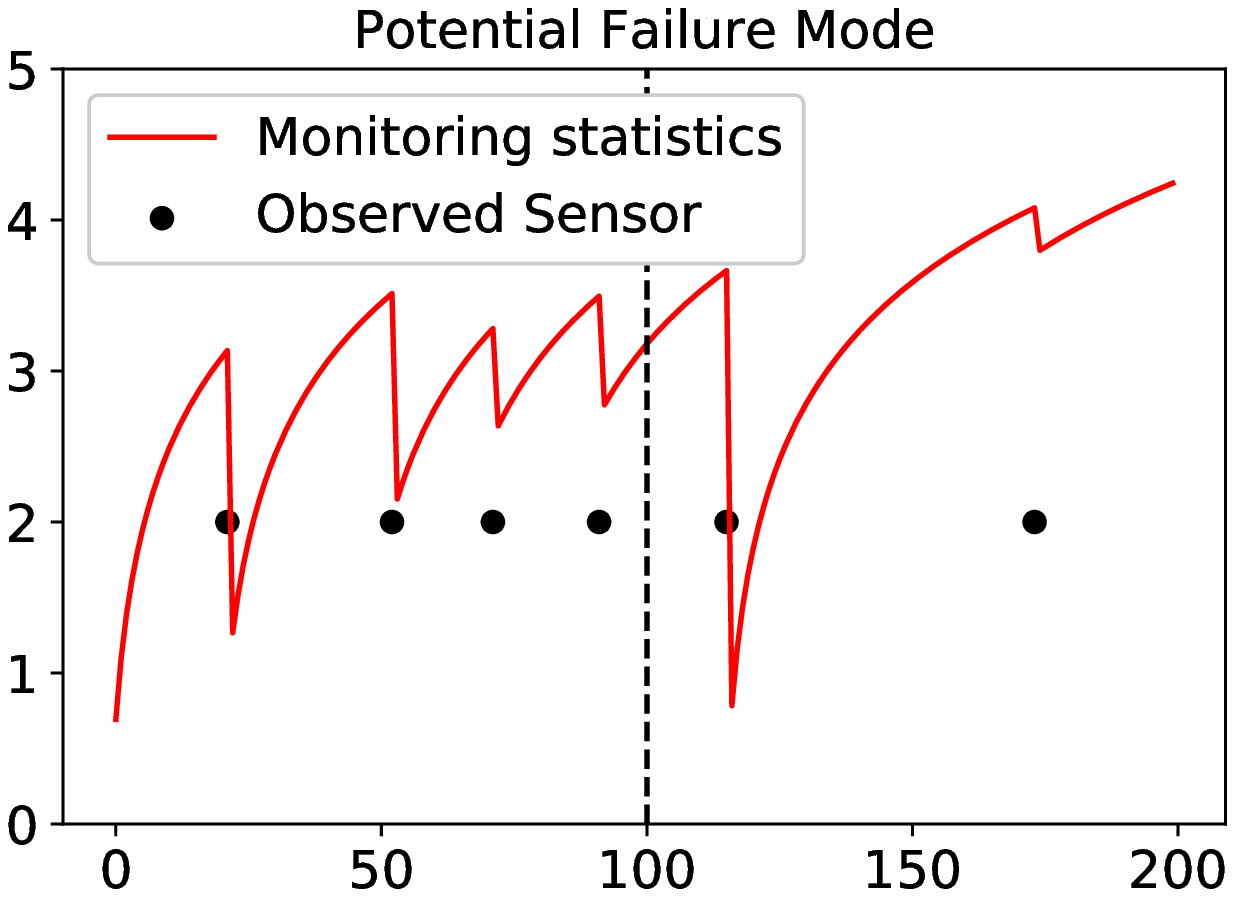}\label{fig:potentialstat}}\subfigure[True Potential Failure Mode ]{\includegraphics[width=0.4\textwidth,height=0.3\textwidth]{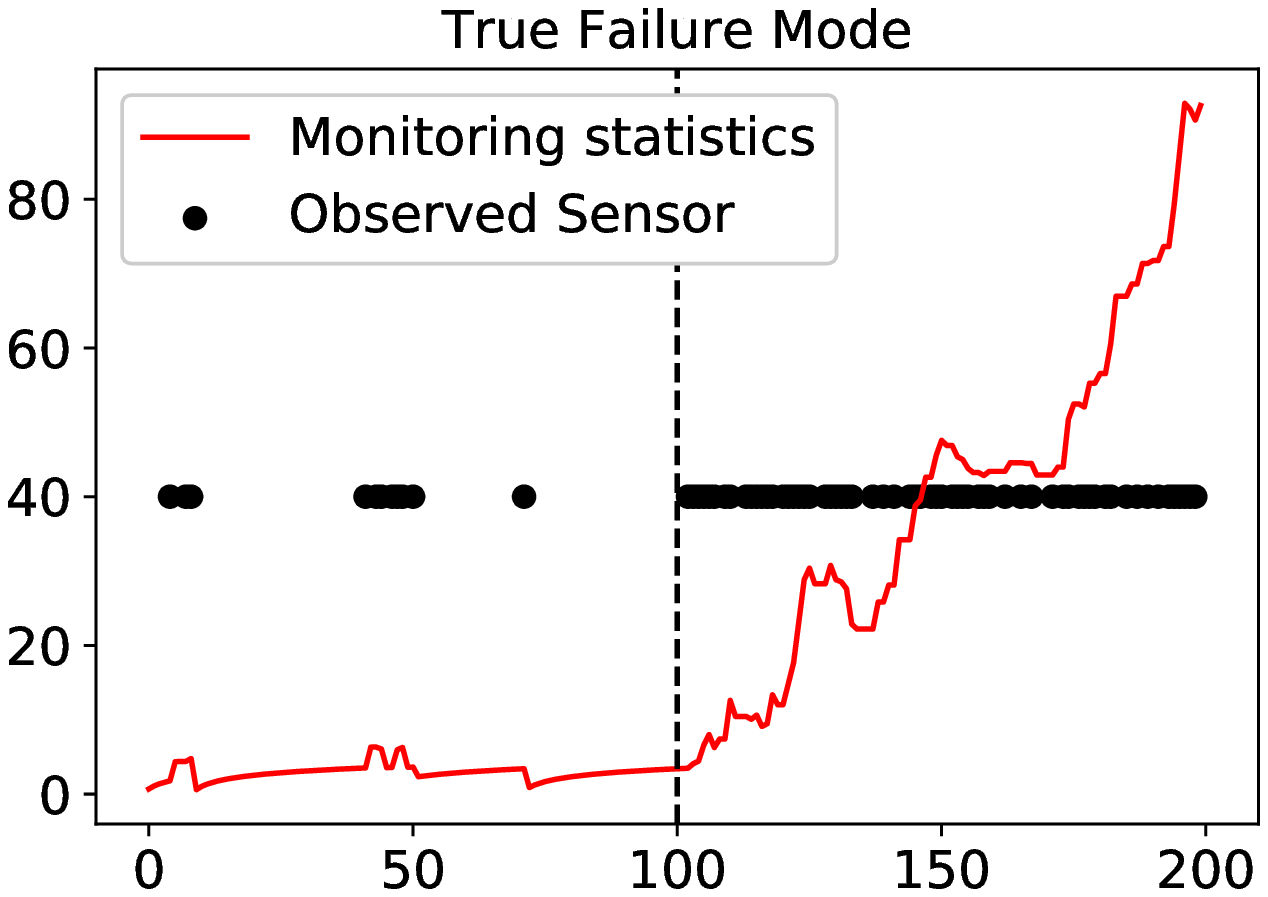}\label{fig:truestat}}
    \caption{SR statistics for the True Failure Mode. The change happens at time $t=100$. Left figure shows the monitoring statistics for the potential failure mode, where the monitoring statistics is small. Right figure shows the monitoring statistics for the true failure mode and the monitoring statistics increase dramatically at time $t=100$.}\label{fig:Statistics}
\end{figure}

To understand how the proposed algorithm balances the exploration and exploitation automatically, we would like to plot both the SR statistics for each failure mode (i.e., in red) and when this particular failure mode has observed sensors (i.e., in black dot) for both the potential failure mode (i.e., failure mode doesn't happen in this run) and the true failure mode in Fig.~\ref{fig:Statistics} together. Here, the failure mode with the observed sensors can be defined as that there are observed sensing variables located in the non-zero location of the mean-shift of that particular failure mode.  From Fig.~\ref{fig:Statistics}, it is clear that the statistics for potential is quite small compared to the statistics for the true failure mode. From Fig.~\ref{fig:potentialstat}, we can also observe that the SR statistics will grow naturally if this particular failure mode is not observed. This will encourage the sensing variables to be allocated to this particular failure mode eventually. Furthermore, if the particular failure mode is observed where no change occurs, the monitoring statistics will drop significantly, indicating that this failure mode has dropped significantly. On the other hand, from Fig.~\ref{fig:truestat}, we can clearly see that after time $100$ when the change occurs, the true failure mode statistics will grow significantly as long as that particular failure mode is observed. 

\subsection{Simulation Study for Multiple Failure Modes \label{subsec: simu_muti}}
We further design another simulation study to evaluate the performance of the proposed methods when multiple failure modes happen together. Fig.~\ref{fig:sim_anomaly} illustrates the potential failure modes and generated multiple failure modes.
\begin{figure}
    \centering
        \subfigure[All 50 Non-overlapping potential failure modes]{\includegraphics[width=0.4\textwidth,height=0.4\textwidth]{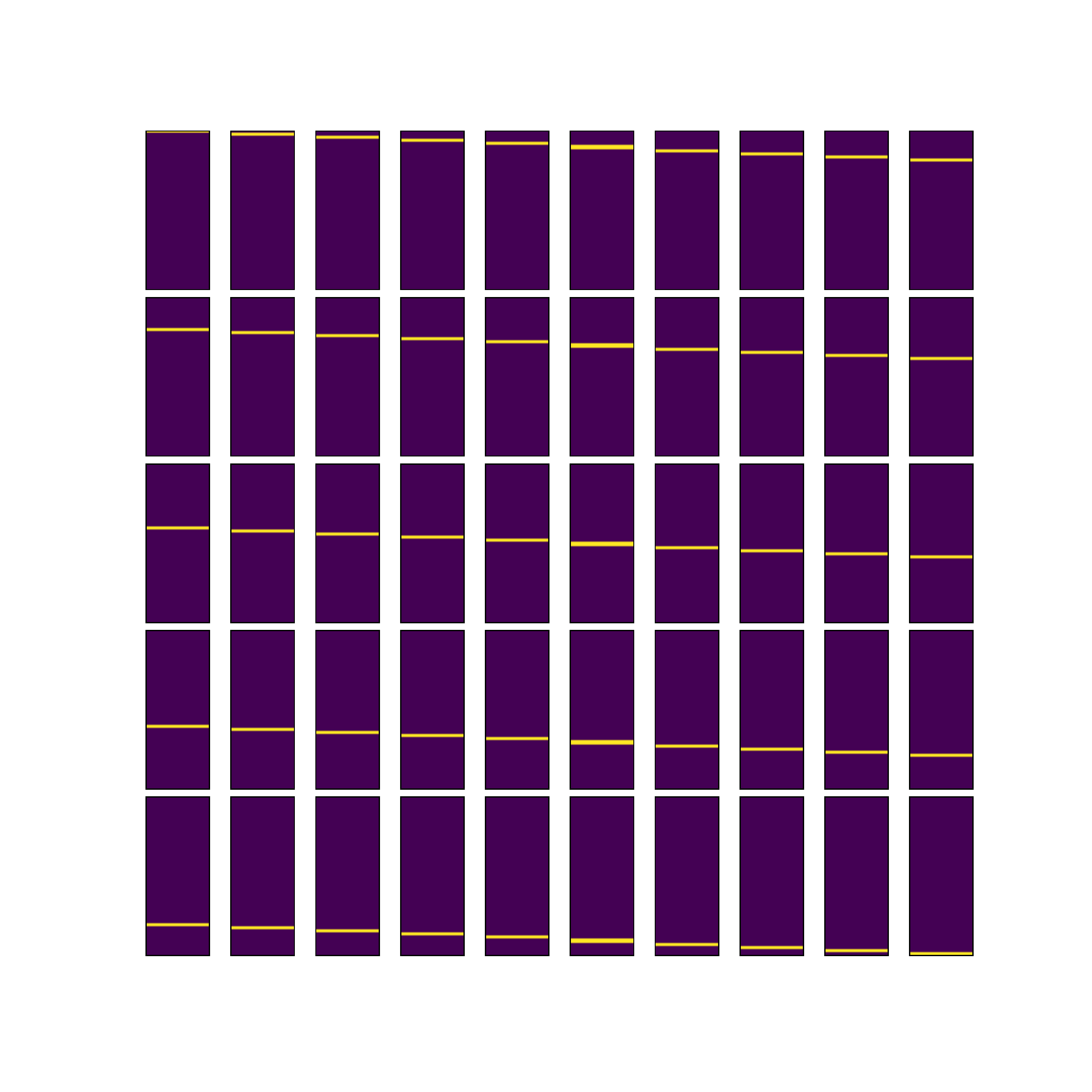} \label{fig:sim_iid_potential}} \subfigure[All 49 overlapping potential failure modes]{\includegraphics[width=0.4\textwidth,height=0.4\textwidth]{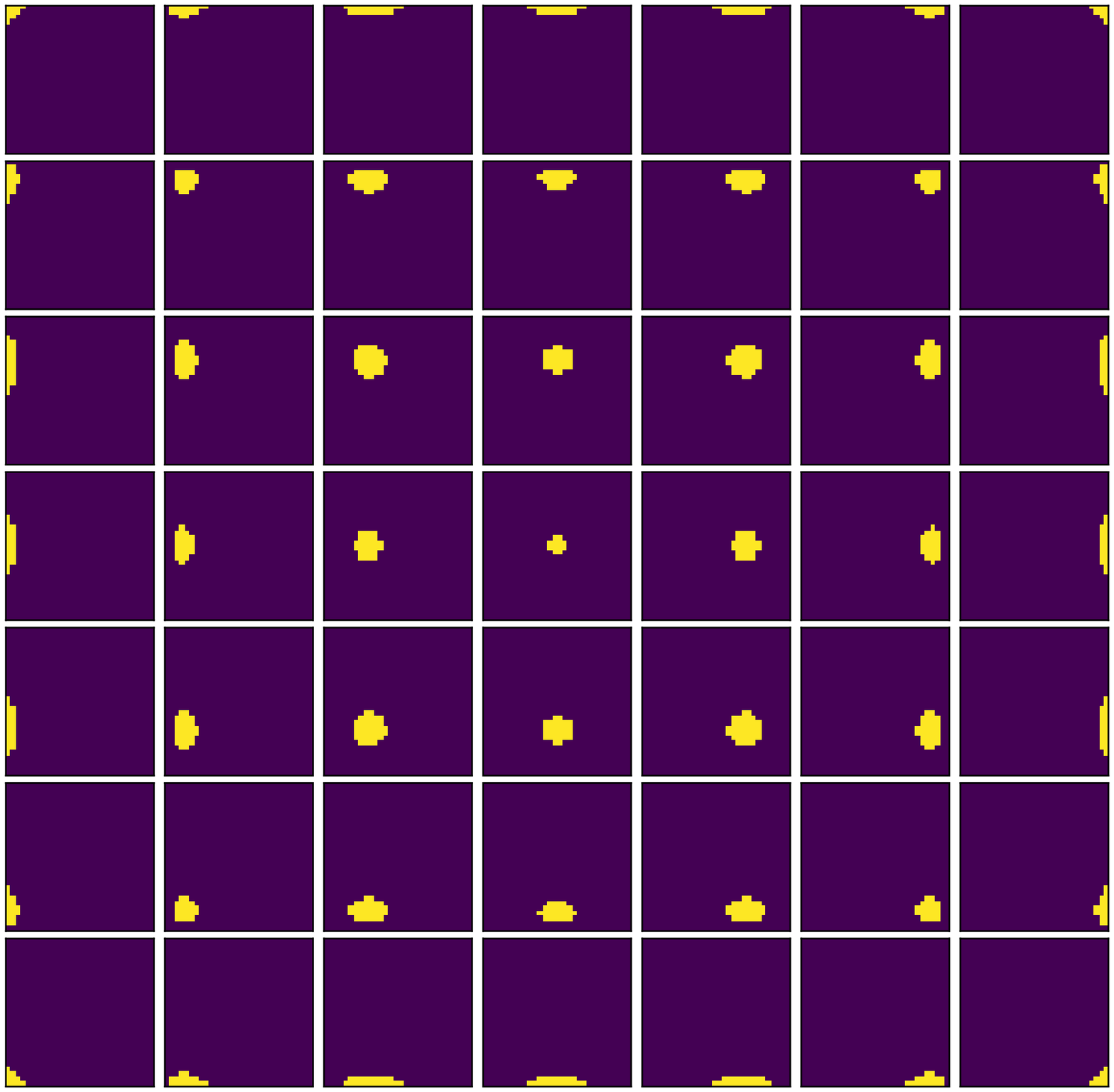}
    \label{fig:sim_spline_potential}}

    \subfigure[Non-overlapping true failure modes]{
  \includegraphics[width=0.4\textwidth,height=0.4\textwidth]{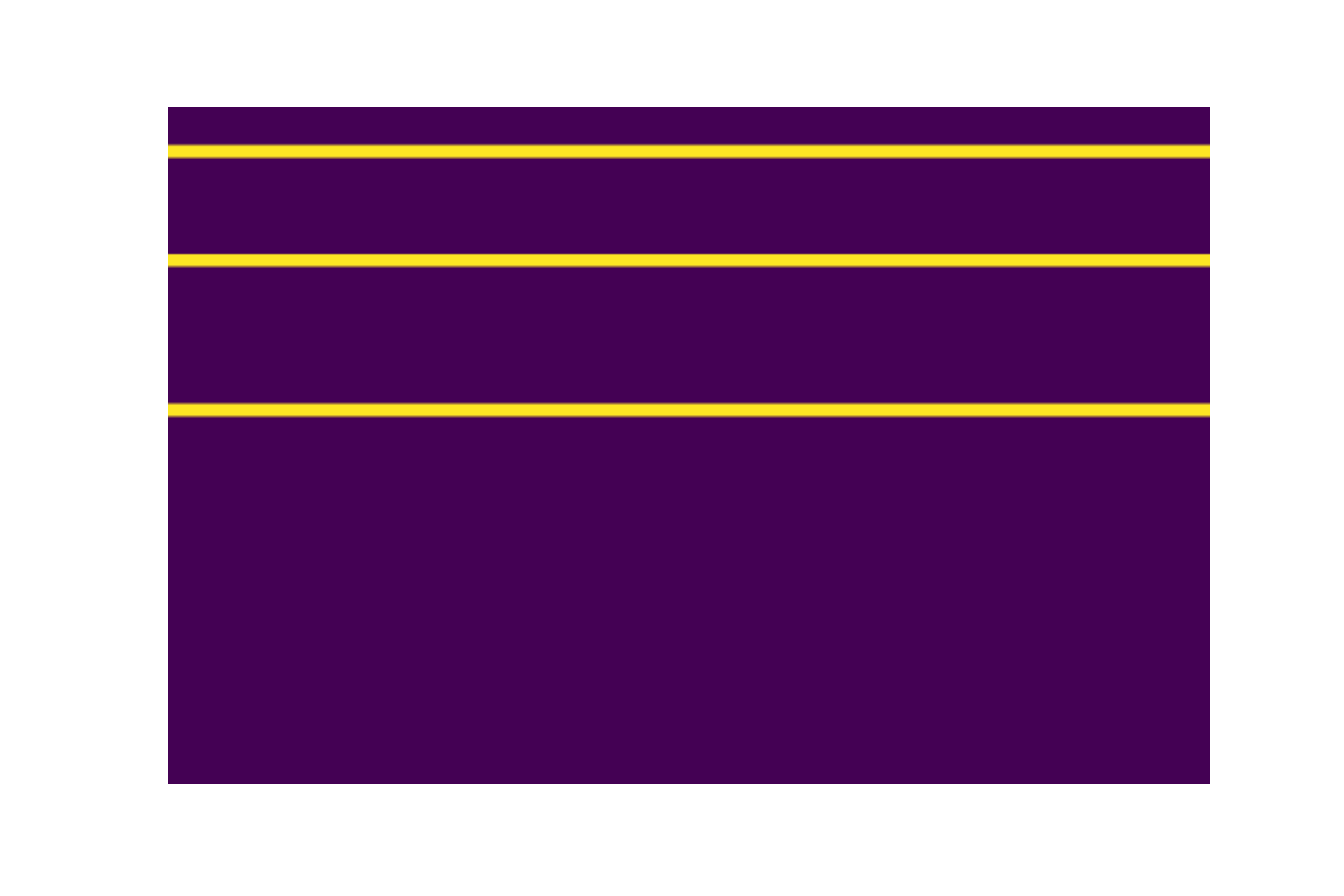}
    \label{fig:sim_iid_true_anomaly}}\subfigure[Overlapping true failure modes]{   \includegraphics[width=0.4\textwidth,height=0.4\textwidth]{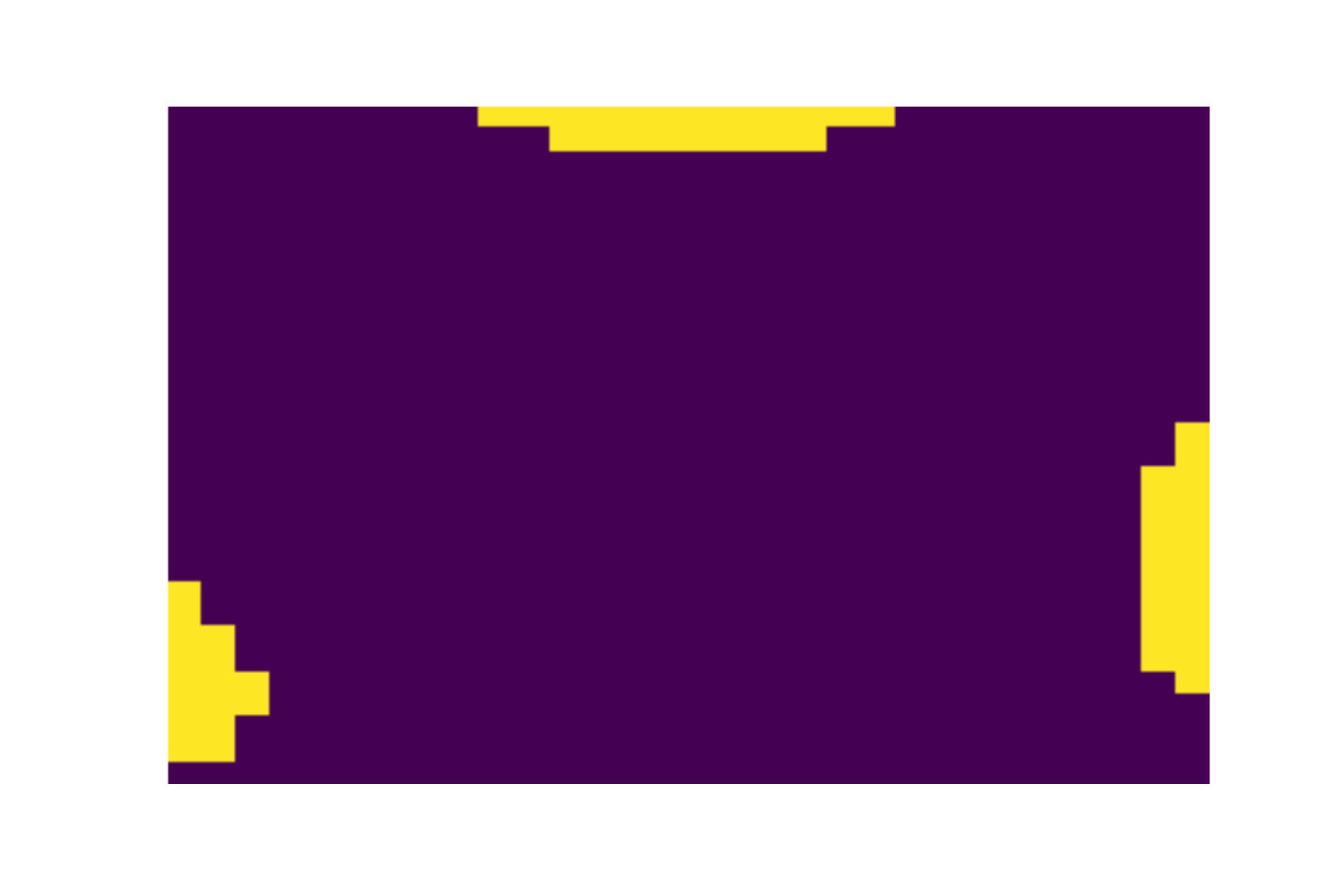}
    \label{fig:sim_spline_true_anomaly}}
    
    \caption{Generated Potential and True Failure Mode Patterns for Overlapping and Non-overlapping Cases}\label{fig:sim_anomaly}
\end{figure}

\begin{table}
\centering \caption{Average Run Length and Failure Mode Isolation Accuracy for multiple failure}
\begin{adjustbox}{width=0.95\textwidth} %

\begin{tabular}{|>{\raggedright}p{2cm}|c|c|c|c|c|c|c|c|c|}
\hline 
\multicolumn{2}{|c|}{\emph{Case} } & \multicolumn{4}{c|}{nonoverlap} & \multicolumn{4}{c|}{overlap}\tabularnewline
\hline 
\multicolumn{2}{|c|}{\emph{Change Magnitude }} & \multicolumn{2}{c|}{$\delta=0.5$} & \multicolumn{2}{c|}{$\delta=0.8$} & \multicolumn{2}{c|}{$\delta=0.5$} & \multicolumn{2}{c|}{$\delta=0.8$}\tabularnewline
\hline 
\multicolumn{2}{|c|}{\emph{Metrics }} & $ARL_1$  & Accuracy  &  $ARL_1$  & Accuracy  &  $ARL_1$  & Accuracy  &  $ARL_1$  & Accuracy \tabularnewline
\hline 
\multicolumn{2}{|c|}{\emph{Oracle }} & 6.05(0.12)  & 0.91(0.29)  & 1.78(0.02)  & 0.93(0.26)  & 5.28(0.12) & 0.99(0.09)  & 1.56(0.02)  & 1.0(0.0) \tabularnewline
\hline 
\multirow{5}{2cm}{\emph{Competing Methods}} & MTSSRP  & \textbf{57.26(1.32) } & \textbf{0.91(0.29) } & \textbf{14.27(0.22) } & \textbf{0.93(0.26) } & \textbf{53.73(1.36) } & \textbf{ 0.93(0.25) } & \textbf{14.38(0.23) } & \textbf{0.97(0.16) }\tabularnewline

\cline{2-10} \cline{3-10} \cline{4-10} \cline{5-10} \cline{6-10} \cline{7-10} \cline{8-10} \cline{9-10} \cline{10-10} 
 & TSSRP  & 77.78(1.26)  & 0.85(0.36)  & 36.32(0.53)  & 0.88(0.26)  & 79.15(1.35)  & 0.89(0.31)  & 32.12(0.54)  & 0.92(0.26) \tabularnewline
 
\cline{2-10} \cline{3-10} \cline{4-10} \cline{5-10} \cline{6-10} \cline{7-10} \cline{8-10} \cline{9-10} \cline{10-10} 
 & TRAS  & 104.18(1.67)  & 0.43(0.50)  & 38.1(0.56)  & 0.91(0.29)  &  102.81(1.71)  & 0.48(0.50)  & 39.34(0.68)  & 0.93(0.25) \tabularnewline
 
\cline{2-10} \cline{3-10} \cline{4-10} \cline{5-10} \cline{6-10} \cline{7-10} \cline{8-10} \cline{9-10} \cline{10-10} 
 & MRandom  & 146.09(1.64)  & 0.59(0.49)  & 64.18(0.82)  & 0.92(0.26)  & 149.29(1.69)  & 0.57(0.50)  & 69.63(1.06)  & 0.98(0.15) \tabularnewline
 
\cline{2-10} \cline{3-10} \cline{4-10} \cline{5-10} \cline{6-10} \cline{7-10} \cline{8-10} \cline{9-10} \cline{10-10} 
 & Random  & 199.96(0.03)  & -  & 198.6(0.29)  & -  & 199.93(0.05)  & -  & 198.44(0.32)  & - \tabularnewline
\hline 
\end{tabular}\label{table: ARLsim} 

\end{adjustbox} 
\end{table}

\begin{figure}
    \centering
    \subfigure[$ARL_1$ for the Non-overlapping Cases]{
\includegraphics[width=0.45\textwidth,height=0.36\textwidth]{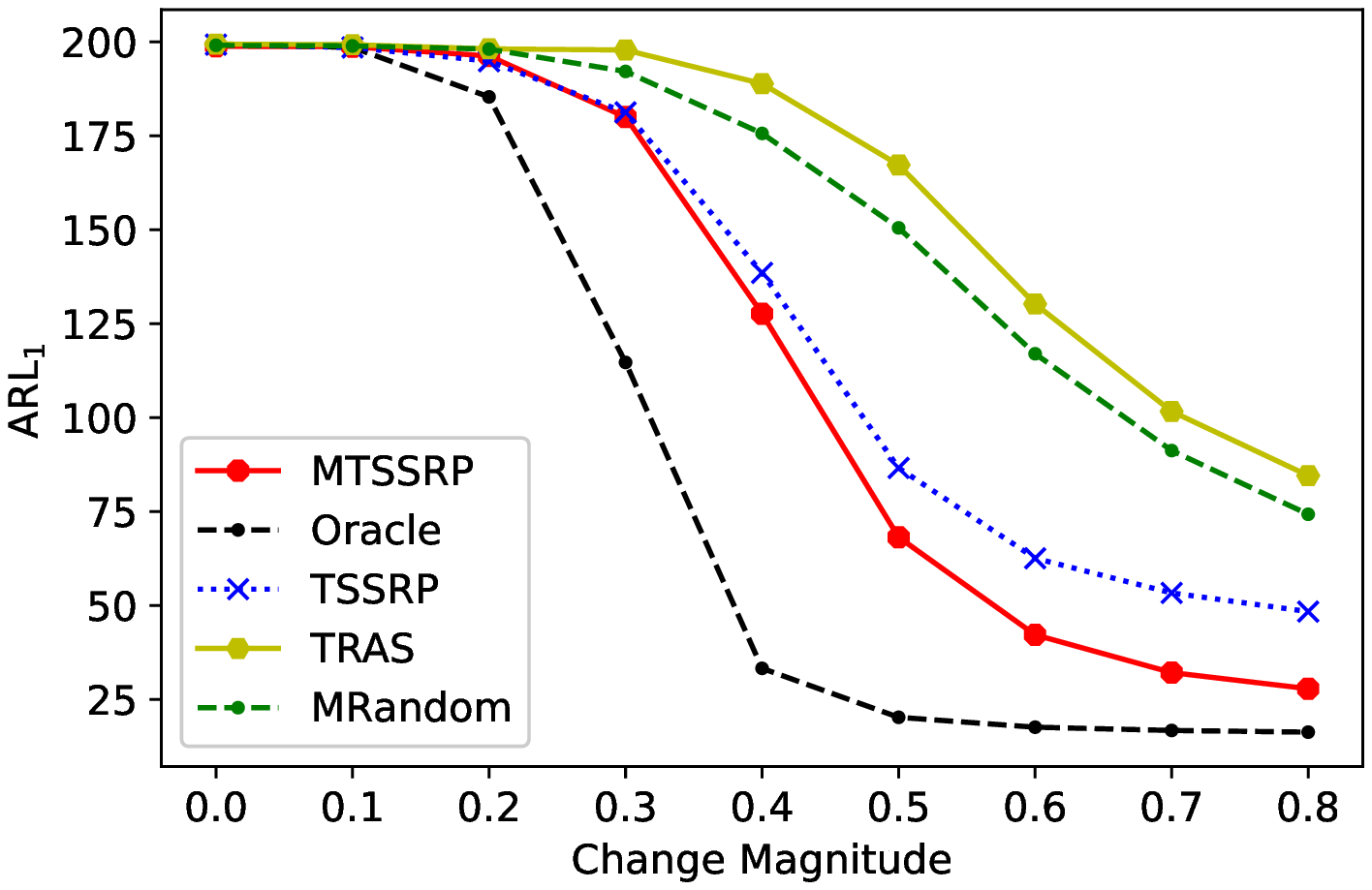}
    \label{fig:ARL1}}
    \subfigure[$ARL_1$ for Overlapping Cases]{
  \includegraphics[width=0.45\textwidth,height=0.36\textwidth]{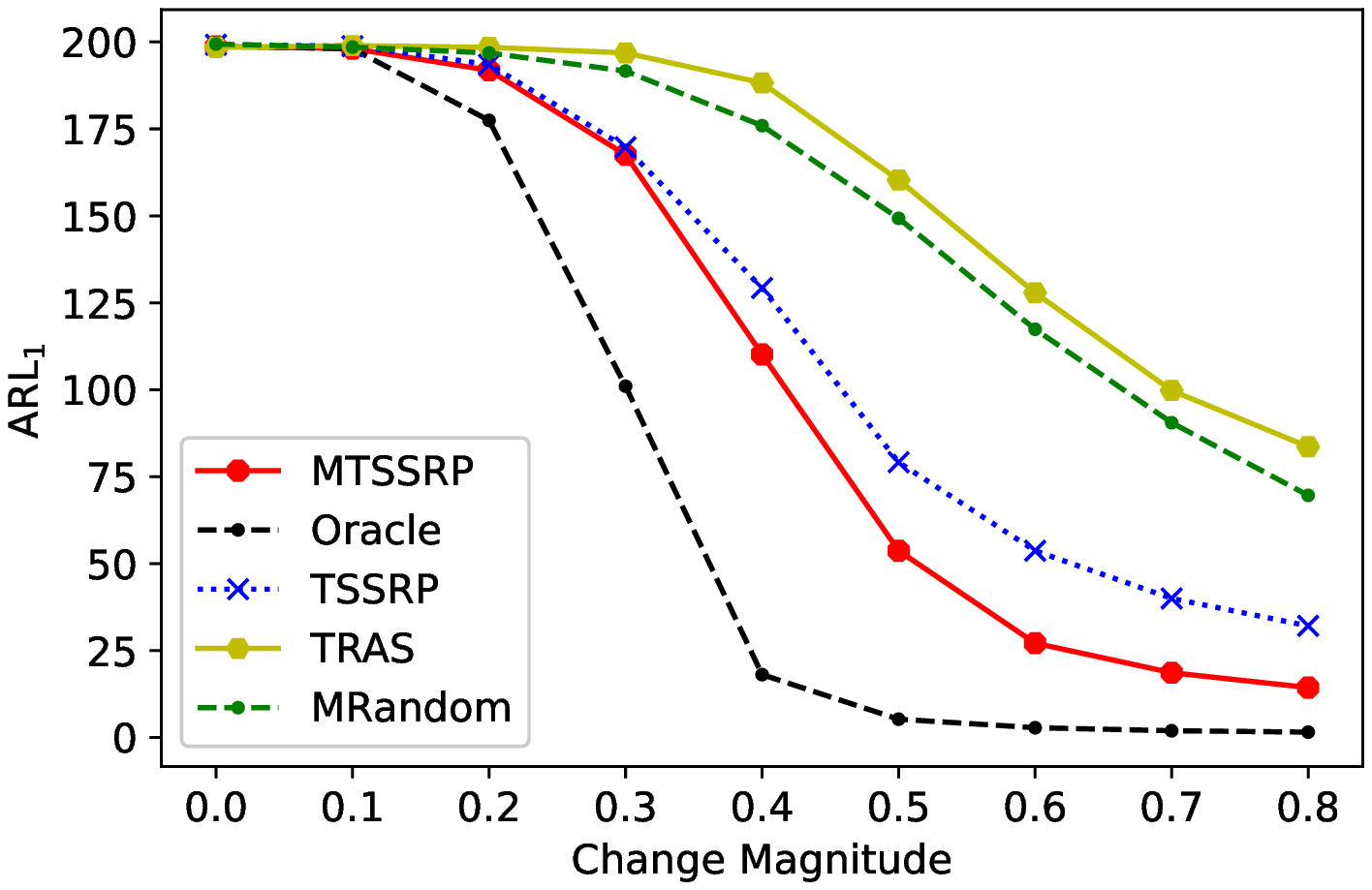}
    \label{fig:ARL2}}
    
    \subfigure[Accuracy for the Non-overlapping Cases]{
\includegraphics[width=0.45\textwidth,height=0.3\textwidth]{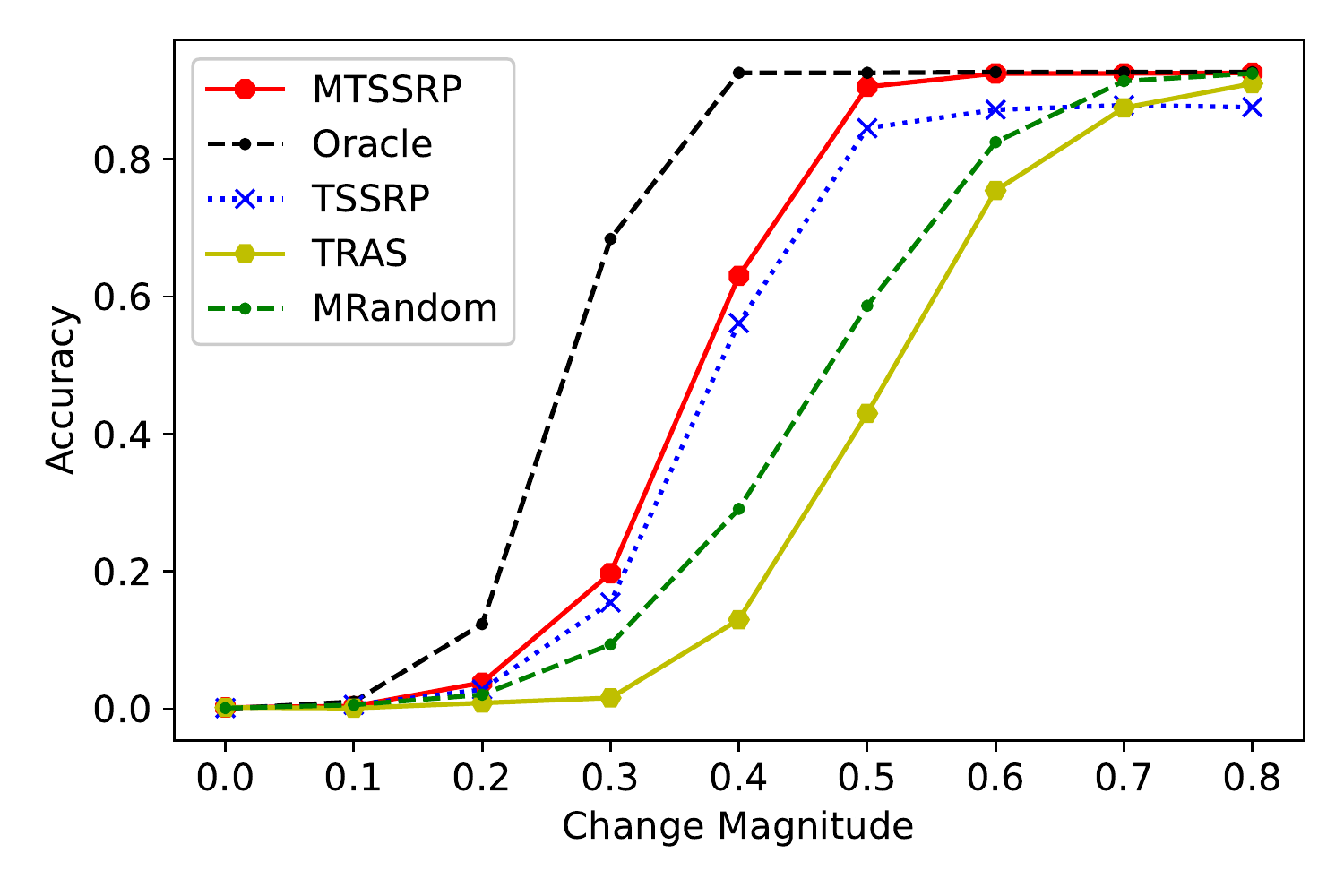}
    \label{fig:accuracy1}}
    \subfigure[Accuracy for Overlapping Cases]{
  \includegraphics[width=0.45\textwidth,height=0.3\textwidth]{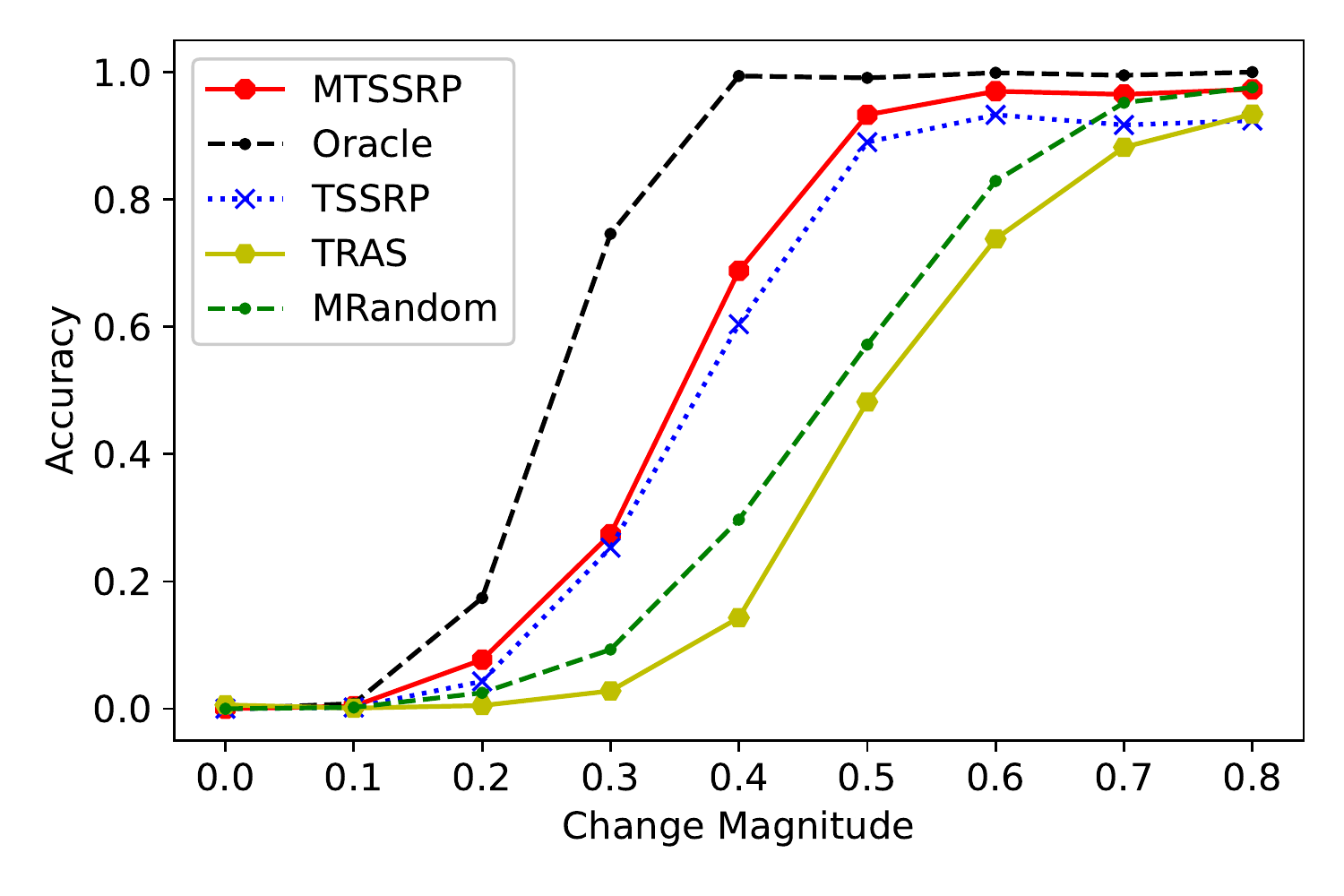}
    \label{fig:accuracy2}}
    \caption{Out-of-control Average Run Length ($ARL_1$) and Failure isolation accuracy for Different Change Magnitude $\delta$ for multiple failure modes} \label{fig:ARL_multiple}
\end{figure}

Similar to the experiment in the single failure mode. we will compare the proposed MTSSRP with the following benchmark methods: 1) TSSRP method \citep{zhang2020bandit}. 2) TRAS method \cite{liuAdaptive2015}, where the local CUSUM statistics are used for each individual data stream and later fused together via the Top-R rule. To show the upper-bound and lower-bound performance, we will also add three simple alternatives: 1) Random, where we randomly select $q$ sensors at each time step with the top-r statistics by monitoring each sensor individually. 2) Oracle, where we have not only access to all the data streams but also the failure mode distribution information using the monitoring statistics as MTSSRP. 3) MRandom, where we apply the same monitoring statistics as MTSSRP, which considers the failure mode distribution information in the monitoring statistics, but we randomly select the sensors at each time step. We evaluate the proposed method with two metrics which are detection delay and failure isolation accuracy. Here, we will compute the $ARL_1$ for different change magnitudes ($\delta$ = $0.5$, $0.8$) in Table \ref{table: ARLsim}. We also compared the $ARL_1$ and Isolation Accuracy from different magnitude $\delta$ ($0.1$ to $0.8$) in Fig.~\ref{fig:ARL_multiple}

\begin{figure}
    \centering
    \subfigure[IC Sampling Point Distribution for Non-overlapping Failure Modes]{
\includegraphics[width=0.3\textwidth,height=0.3\textwidth]{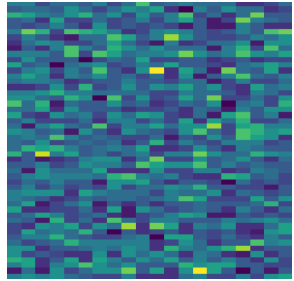}
    \label{fig:pointICnon}} \quad
    \subfigure[OC Sampling Point Distribution for Non-overlapping Failure Modes]{
    \hspace*{15pt}    \includegraphics[width=0.3\textwidth,height=0.3\textwidth]{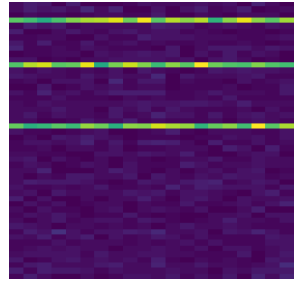}
    \label{fig:pointOCnon}}
    \subfigure[IC Sampling Point Distribution for Overlapping Failure Modes]{
    \hspace*{15pt}
    \includegraphics[width=0.3\textwidth,height=0.3\textwidth]{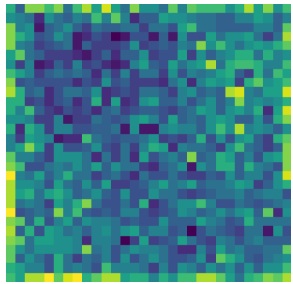}
    \hspace*{15pt}
    \label{fig:pointICovl}} \quad\subfigure[OC Sampling Point Distribution for Overlapping Failure Modes]{
    \hspace*{15pt}
    \includegraphics[width=0.3\textwidth,height=0.3\textwidth]{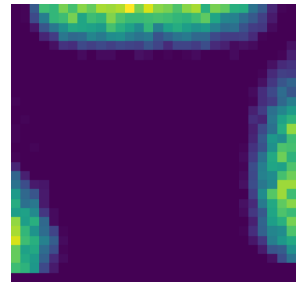}
    \hspace*{15pt}
    \label{fig:pointOCovl}}
    \caption{Examples of Simulated Data and Sampled Point Distribution for Both the Overlapping and Non-overlapping Failure Modes; It is clear that the OC sampling point distributions are the same as the true failure mode patterns.} \label{fig:samplepatern}
\end{figure}

We then show the sampling point distributions before the change and after the change for both scenarios. To visualize the sampling distribution better, we have aggregated 1000 generate sampled IC data and OC data in Fig.~\ref{fig:pointICnon} and Fig.~\ref{fig:pointOCnon}, respectively for the non-overlapping failure modes and Fig.~\ref{fig:pointICovl} and Fig.~\ref{fig:pointOCovl} for the overlapping failure modes. It is clear that before the change, the sampling distribution is pretty random. After the change, most of the points will gather around the three true failure modes. 

\begin{figure}
    \centering
    \subfigure[Non-overlapping case]{\includegraphics[width=0.49\textwidth]{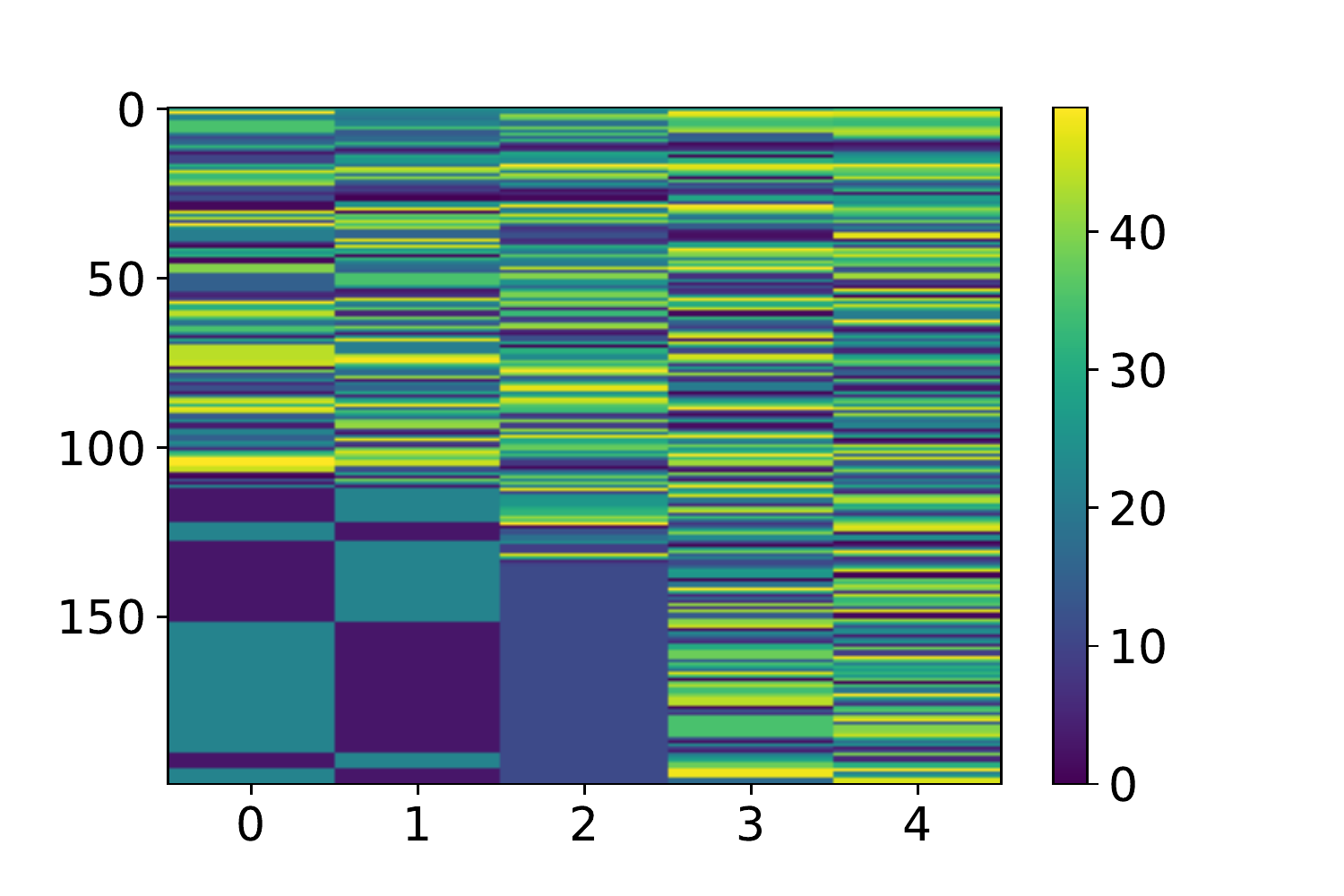}\label{fig:failureA}}\subfigure[Overlapping case]{\includegraphics[width=0.49\textwidth]{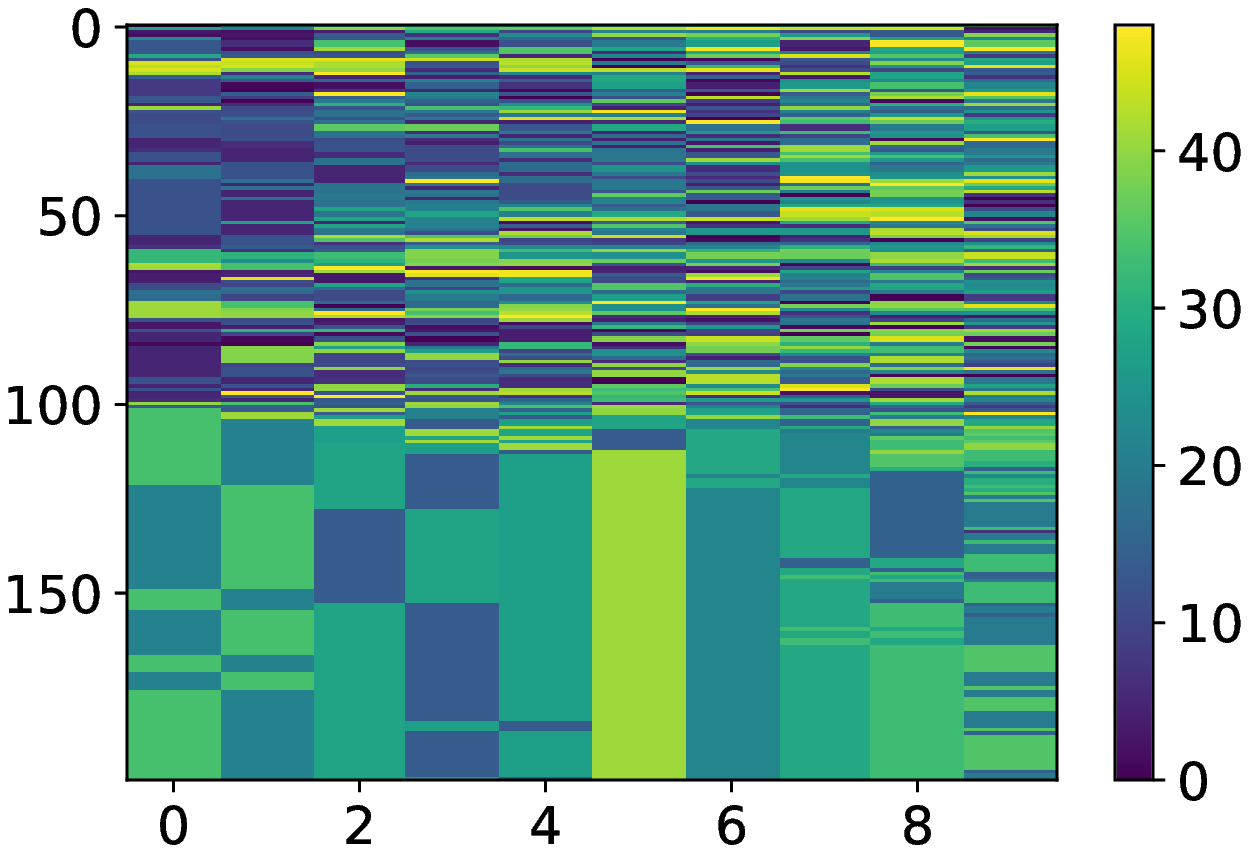}\label{fig:failureB}}
    \caption{Failure Mode Selection history of MTSSRP. Here, X-axis refers to the ranking of the failure mode, and Y-axis refers to the time. All the changes happen at time 100. }\label{fig:failuremode}
\end{figure}

Finally, we showed the heatmap of the identified top failure modes from MTSSRP method at $\delta=4$ in Fig.~\ref{fig:failuremode}. It shows the top most likely failure modes  (columns) at a different time (rows) from our proposed algorithm. The actual change point of the data is at $t=100$. Column $0$ shows that we identify a failure pattern at around time $100$, which is consistent with the change time. Prior to time $100$, the failure mode patterns are actually quite random. Interestingly, we find that after the algorithm finds the first failure pattern, the algorithm will continue to search for other underlying failure patterns. At around time 120, we further detect the other two failure patterns in our experiments in both overlapping and non-overlapping cases. Another interesting behavior is that the other failure modes (except the top 3) are quite random in the non-overlapping case but not as random in the overlapping case. The reason is that in the non-overlapping failure modes, given that potential failure modes have no overlapping with the true failure modes, there are no particular orders on the potential failure modes. However, in the overlapping case, besides the 3 true failure modes, there are also some potential failure modes overlapped with the true failure modes, which will be selected. 
In conclusion, this behavior shows that the proposed algorithm is able to detect the true failure modes, as indicated by the Theorem \ref{Limit}. 

\section{Case Study}  \label{sec: casestudy}
In this section, we will evaluate the proposed MTSSRP algorithm in the laser powder bed fusion process monitoring. We will evaluate the performance of the MTSSRP and compare it with the state-of-the-art benchmark methods.


\begin{figure}
    \centering
    \includegraphics[width=0.5\textwidth]{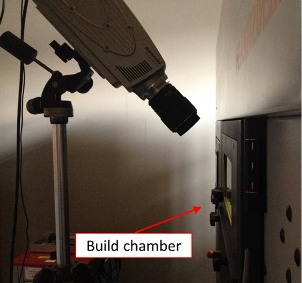}
    \caption{Hot-spots Detection in LPBF}
    \label{fig:hotspotsetup}
\end{figure}

\subsection{Hot-spots Detection in Laser Powder Bed Fusion Process} \label{subsec: additive}

Here, we will implement the proposed algorithm into the hot-spots detection in the process monitoring in the Laser Powder Bed Fusion (LPBF) process. A $300$ fps video sequence was acquired during the realization of one layer of the part by using the setup shown in Fig.~\ref{fig:hotspotsetup}, which consists of a thermal camera mounting outside the LPBF chamber monitoring the hot-spots events. The observed image is of size $121\times71$ pixels. 
Previous studies showed that the occurrence of local over-heating conditions might yield geometrical distortions \citep{yan2020real,colosimo2018spatially}. 
The hot-spots caused by the formation of solidified balls will cause the local heat accumulation and inflate from one layer to another. Therefore, the overall goal of this study is to detect such hot-spots quickly. For more details about the setup of this experiment and some preliminary works related to this dataset, please refer to \citep{grasso2017process,colosimo2018spatially,yan2020real}. The dataset is also publicly available at \url{http://doi.org/10.6084/m9.figshare.7092863}.

\begin{figure}[h!] 
\centering
\includegraphics[width=0.9\textwidth]{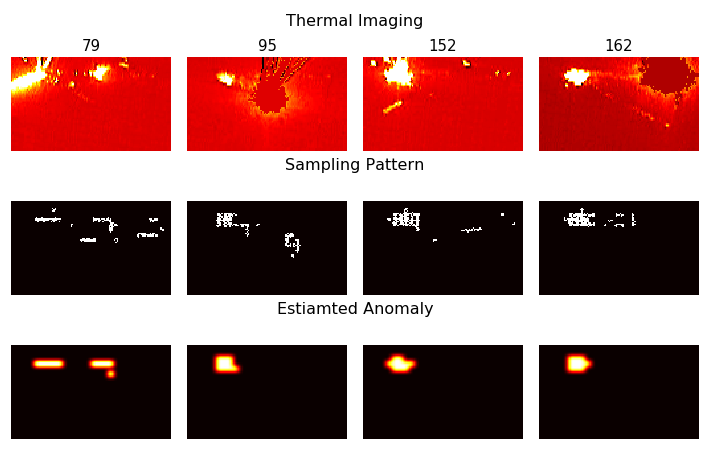}
\caption{Original Thermal Images, Sampling Patterns and Detected Failure Modes}
\label{am}
\end{figure}

In this example, it is not easy to obtain the failure mode data beforehand, and therefore, we rely on domain knowledge to define the failure mode distribution. First, we know that the hot-spots must be in the scanning path. Second, we know that the hot-spots must be locally clustered. Therefore, we define the failure modes as each individual B-spline basis overlapped with the printing regions. 
In this dataset, there are four different events starting from 77, 94, 150, and 162. From Table \ref{LPBFdetect}, the proposed algorithm can detect the change at time 79, 95, 152, and 162 with only 200 sensing variables out of $8591$ sensing variables. In comparison, TSSRP can detect all four events but with a much larger delay. However, TRAS can only detect Event 4, which fails to detect the first three events. 
The original image frame, the sampling patterns, and the selected failure modes at the detected time for these four events are shown in Fig.~\ref{am}. 
From Fig.~\ref{am}, we can observe that the algorithm can quickly converge the sampled points to the true hot-spots location at the upper left corner.

\begin{table}[h!]
\caption{Detection Time in the LPBF Process}

\centering %
\begin{tabular}{|>{\raggedright}p{2cm}|c|c|c|c|c|}
\hline 
\multicolumn{1}{|>{\raggedright}p{2cm}}{} &  & \multicolumn{4}{c|}{Time of first signal}\tabularnewline
\hline 
\multicolumn{2}{|c|}{\emph{Event Times}} & Event 1  & Event 2  & Event 3  & Event 4 \tabularnewline
\hline 
\multicolumn{2}{|c|}{\emph{Actual Change Time}} & 77  & 94  & 150  & 162 \tabularnewline
\hline 
\multirow{3}{2cm}{\emph{Competing Methods}} & MTSSRP  & \textbf{79}  & \textbf{95}  & \textbf{152} & \textbf{162 }\tabularnewline
\cline{2-6} \cline{3-6} \cline{4-6} \cline{5-6} \cline{6-6} 
 & TSSRP  & 80  & 99  & 156  & 164 \tabularnewline
\cline{2-6} \cline{3-6} \cline{4-6} \cline{5-6} \cline{6-6} 
 & TRAS  & -  & -  & -  & 165 \tabularnewline
\hline 
\end{tabular}

\label{LPBFdetect}
\end{table}

\subsection{Tonnage Signal Monitoring}
\begin{figure}
    \centering
    \includegraphics{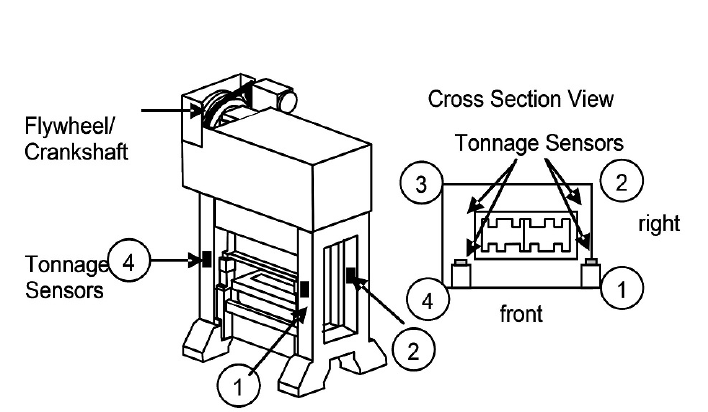}
    \caption{Tonnage Signal Monitoring}
    \label{fig:tonnagesetup}
\end{figure}

We will evaluate the proposed methodology to monitor the tonnage signals collected in a multi-operation forging process, where four strain gauge sensors  on four columns of the forging machine measure the exerted tonnage force of the press uprights as shown in Fig.~\ref{fig:tonnagesetup}. This results in the tonnage profiles in each cycle of operation. 

\begin{figure} 
    \centering
    \includegraphics[width=0.5\textwidth]{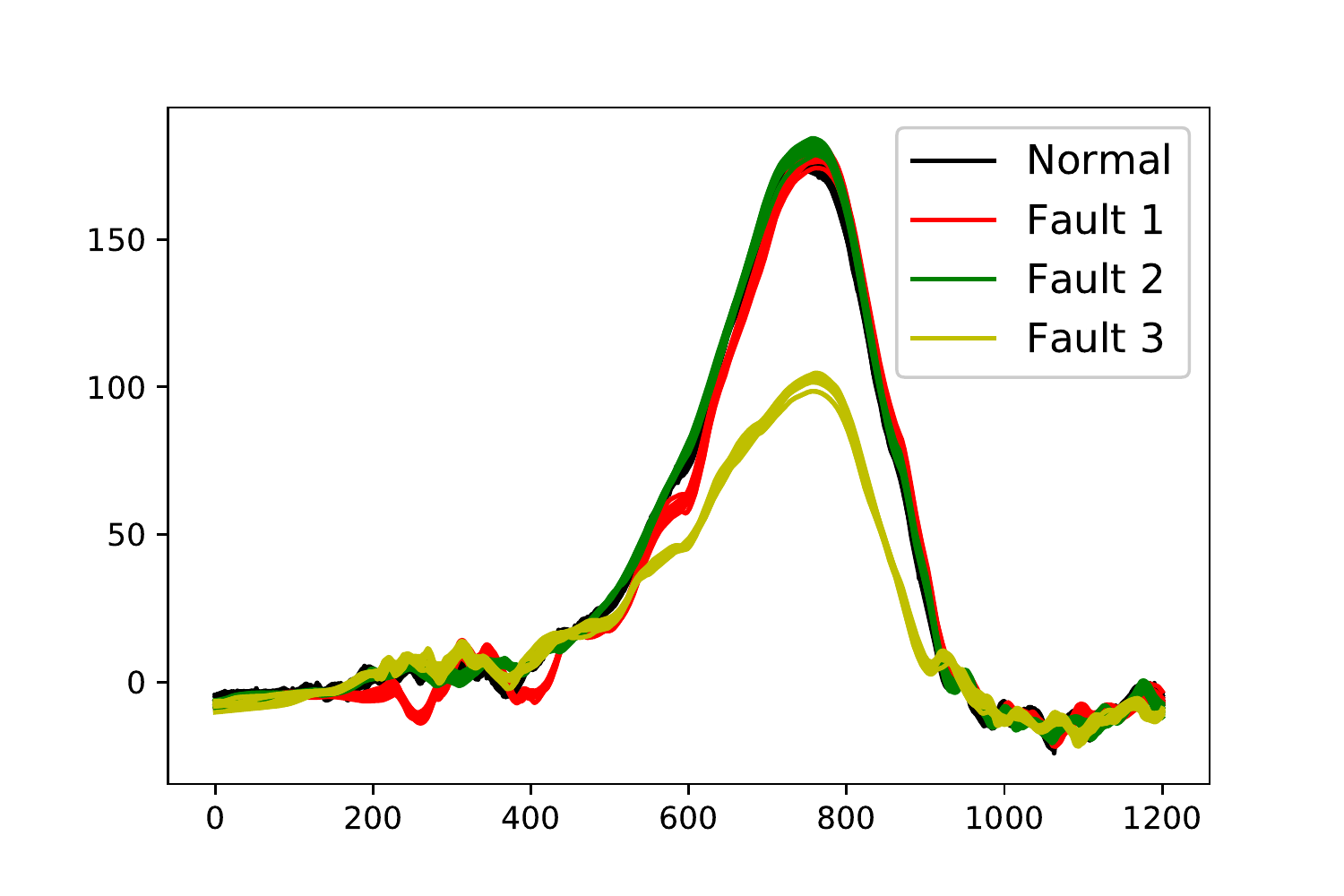}
    \caption{Tonnage Data}
    \label{fig:tonnage}
\end{figure}

The data contains 305 in-control profiles collected under the normal production condition and each abnormal production condition have 68 out-of-control profiles for three different failure modes. 10 samples from each failure mode is shown in Fig.~\ref{fig:tonnage}. The four channel tonnage profiles results in 4804 dimensions in total. 

\begin{figure}
    \centering
    \subfigure[Change of Failure mode 1]{\includegraphics[width=0.33\textwidth]{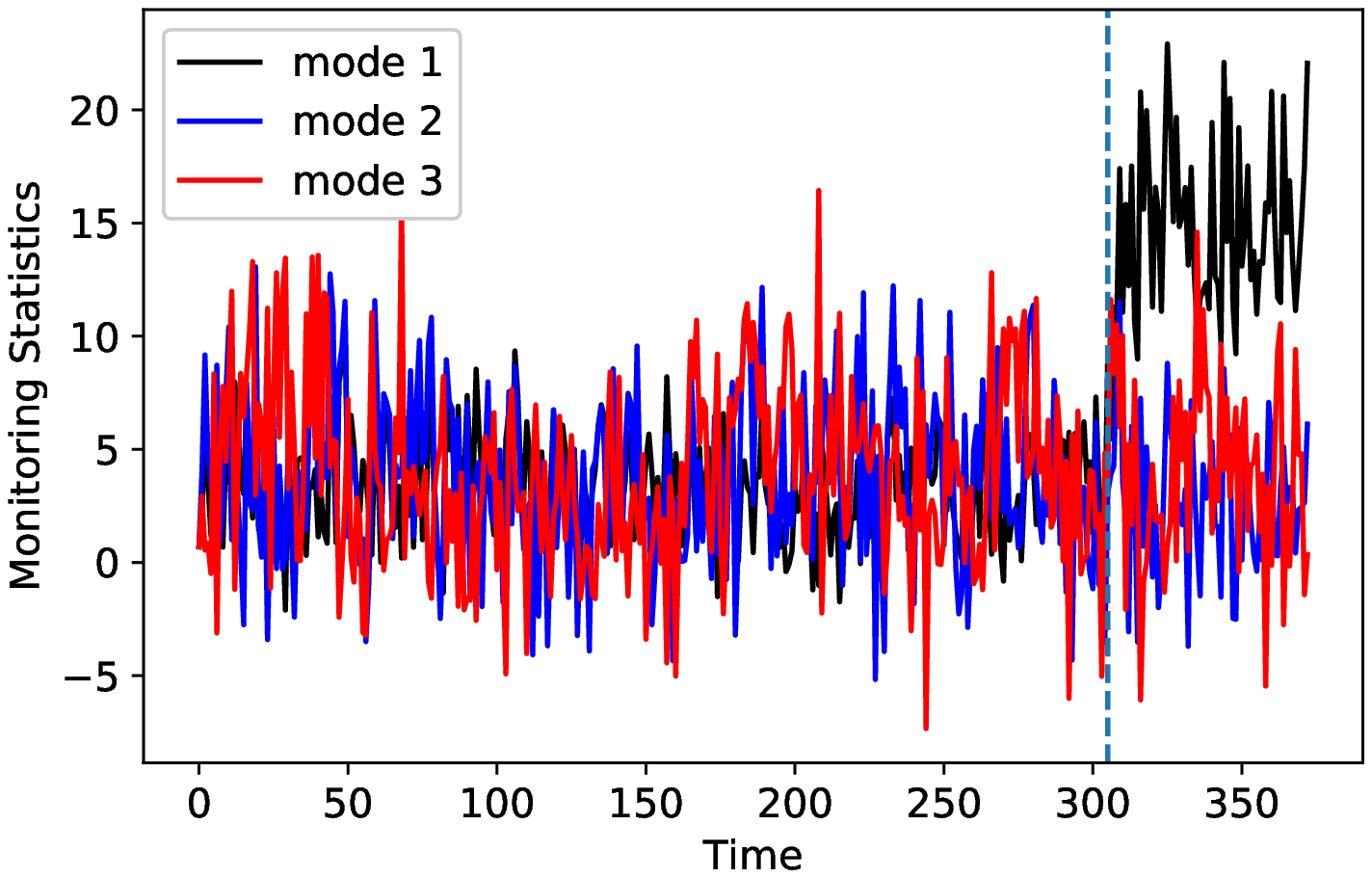}\label{tonnage_failure_1}}\subfigure[Change of failure mode 2]{\includegraphics[width=0.33\textwidth]{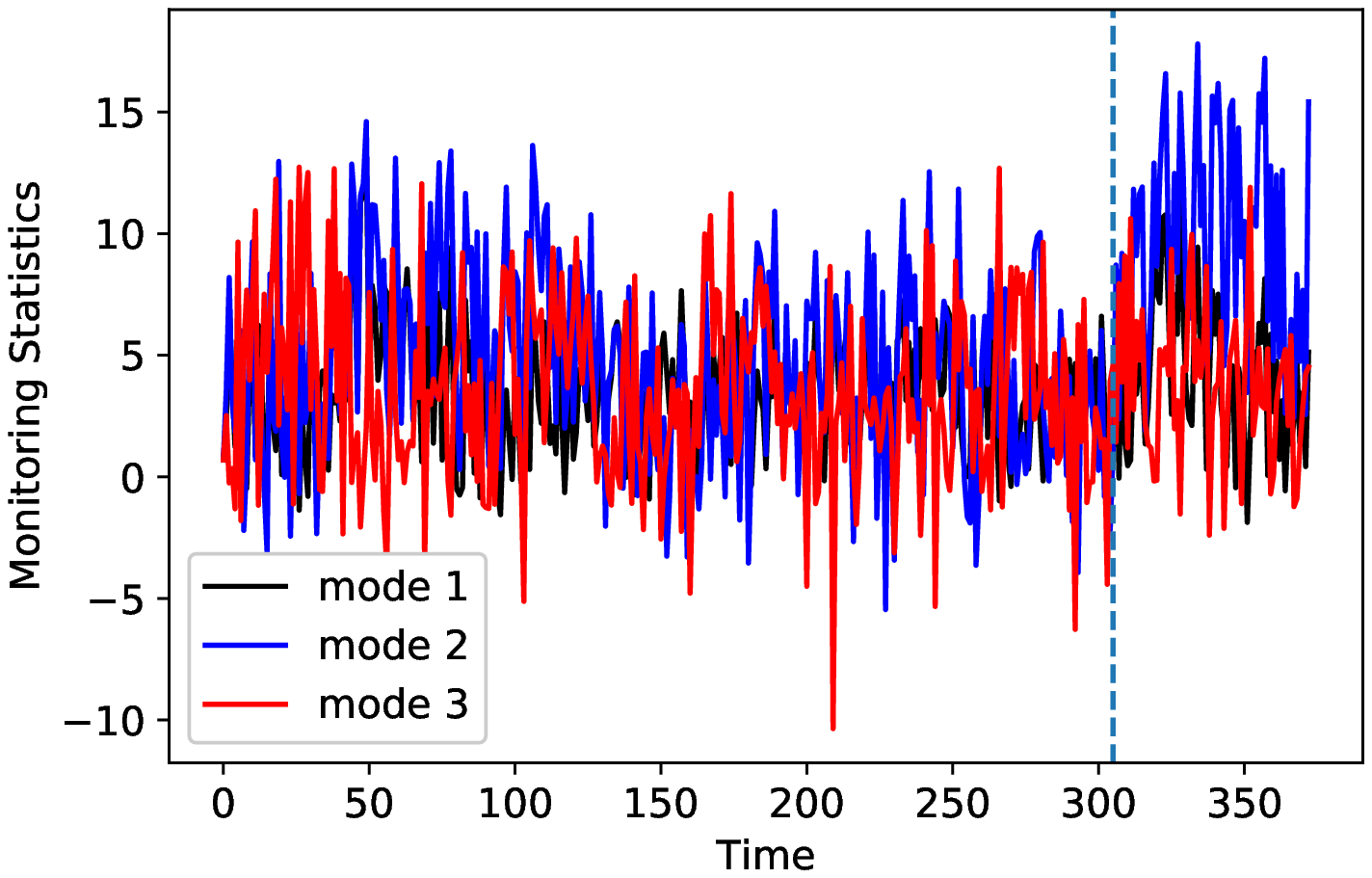}\label{tonnage_failure_2}}\subfigure[Change of failure mode 3 ]{\includegraphics[width=0.33\textwidth]{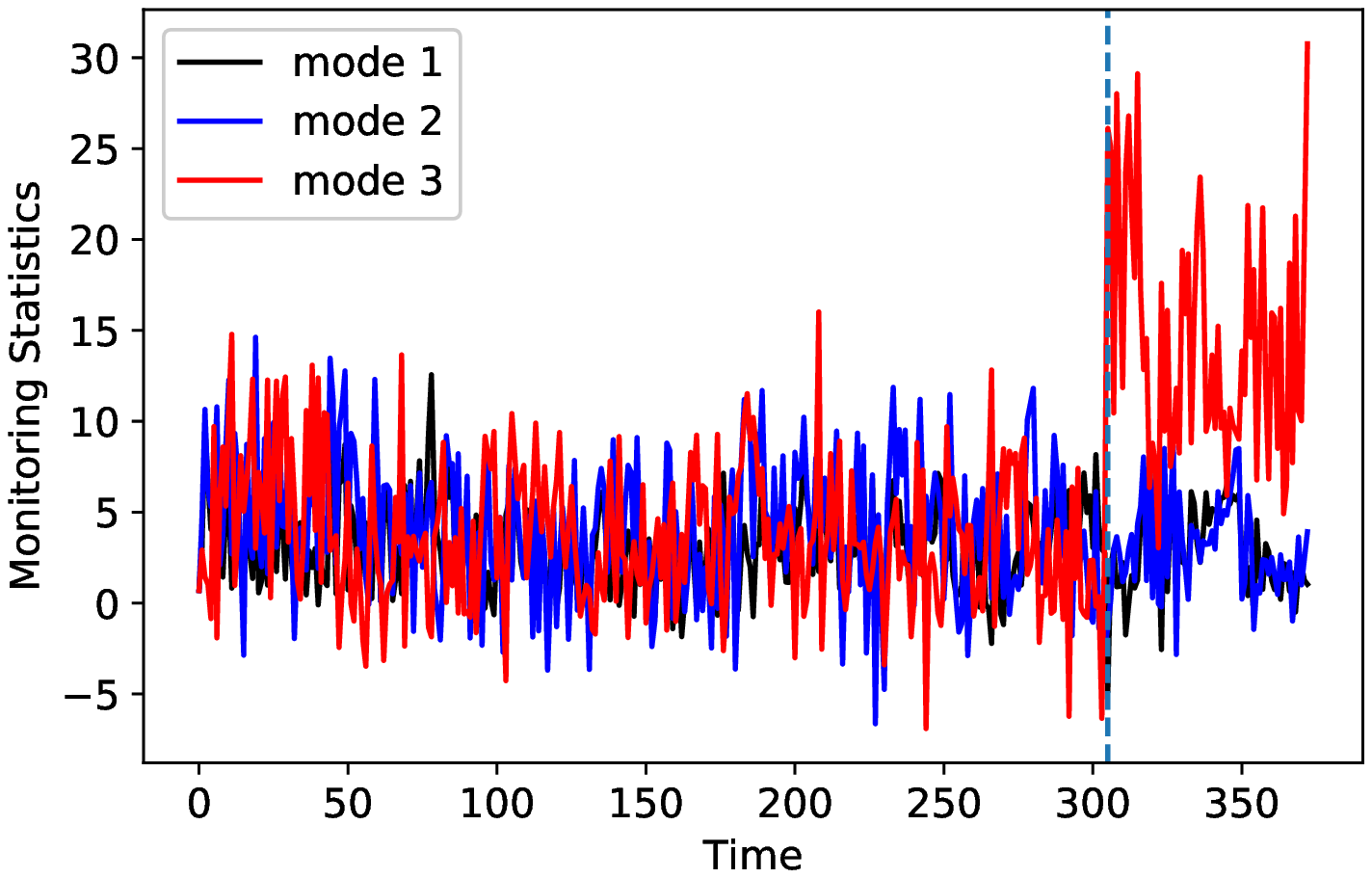}\label{tonnage_failure_3}}
    \caption{Monitoring Statistics for Each Failure Mode. The monitoring statistics of the true failure mode increase significantly after the changed time $t=305$. 
    }\label{fig:resulttonnage}
\end{figure}

In our experiment, since we do not have the prior knowledge about the failure mode,. Therefore, we utilize 20\% of the samples for training of the failure modes by assuming that the data under each failure mode follows the Gaussian distribution with diagonal covariance matrix approximation. We only select 500 out of 4804 data streams for the adaptive sampling. Here, we have conducted a simulation study where the normal samples are randomly drawn from the $305$ in-control profiles. For out-of-control profiles, we have conducted three scenarios, where each scenarios the out-of-control profiles are randomly sampled from 68 out-of-control profiles with replacement. For all the methods, we selected the threshold as the $95\%$ for the IC samples and compare the out-of-control ARL (denoted as $ARL_1$) with $500$ replications for all this three scenarios with the proposed MTSSRP and benchmark methods TSSRP and TRAS. 
The result is shown in Table \ref{tonnagetable}. From Table  \ref{tonnagetable}, we can conclude that failure mode 3 is fairly easy to detect and all methods can achieve $ARL_1$ around $1$. For failure mode 1, it is more similar to the normal dataset. The proposed MTSSRP can achieve $ARL_1$ of $1.96$, which is at least half of the  $ARL_1$ of other methods such as TSSRP and TRAS. Finally, the most challenging case is the failure mode 2, where the difference of failure mode 2 and the normal data is almost neglectable, as seen from Fig.~\ref{fig:tonnage}. The proposed MTSSRP can achieve  $ARL_1$ around $7.46$, which is much smaller than the   $ARL_1$ of TSSRP and TRAS. In Fig.~\ref{fig:resulttonnage}, the monitoring statistics for three failure production conditions have been shown and our method is able to identified the true failure mode correctly. 

\begin{table}
\caption{Out-of-control Average Run Length for Tonnage Signals}

\centering%
\begin{tabular}{|>{\centering}p{2cm}|c|c|c|c|}
\hline 
\multicolumn{2}{|>{\centering}p{2cm}|}{} & \multicolumn{3}{c|}{Out-of-control Average Run Length}\tabularnewline
\hline 
\multicolumn{2}{|c|}{\emph{Failure Modes}} & Mode 1  & Mode 2  & Mode 3\tabularnewline
\hline 
\multirow{3}{2cm}{\emph{Competing Methods}} & MTSSRP  & \textbf{1.96(0.05)}  & \textbf{7.46(0.26)} & \textbf{1.02(0.01)}\tabularnewline
\cline{2-5} \cline{3-5} \cline{4-5} \cline{5-5} 
 & TSSRP  & 4.2(0.15)  & 28.35(0.34) & 1.03(0.02)\tabularnewline
\cline{2-5} \cline{3-5} \cline{4-5} \cline{5-5} 
 & TRAS  & 6.0(0.6)  & 45.85(0.44) & 1.05(0.01)\tabularnewline
\hline 
\end{tabular}

\label{tonnagetable}
\end{table}
\subsection{COVID-19 Cases Detection for Hotspot Detetion}

To better understand the COVID-19 status, different types of testing resource is typically distributed into different regions. Centers for Disease Control and Prevention (CDC) has classified the testing for COVID-19 into the following two categories: 1) \textbf{Diagnostic testing} is intended to identify current infection in individuals and is performed when a person has signs or symptoms consistent with COVID-19, or is asymptomatic, but has recently known or suspected exposure to COVID-19. 2) \textbf{Screening tests} are recommended for unvaccinated people to identify those who are asymptomatic and do not have known, suspected, or reported exposure to COVID-19. Screening helps to identify unknown cases so that measures can be taken to prevent further transmission. \cite{covid192022}. 

Overall, the screening test is very useful in randomly distributed test in some underdeveloped areas to identify unknown cases so that measures can be taken to prevent further transmission. However, in the Screening test, the decision maker may have limited sampling resources. Therefore, adaptive sampling techniques can be used to decide which region to sample at each time epoch based on the collected testing results from the previous iterations. 

\begin{figure}[h!] 
\centering
\includegraphics[width=0.7\textwidth]{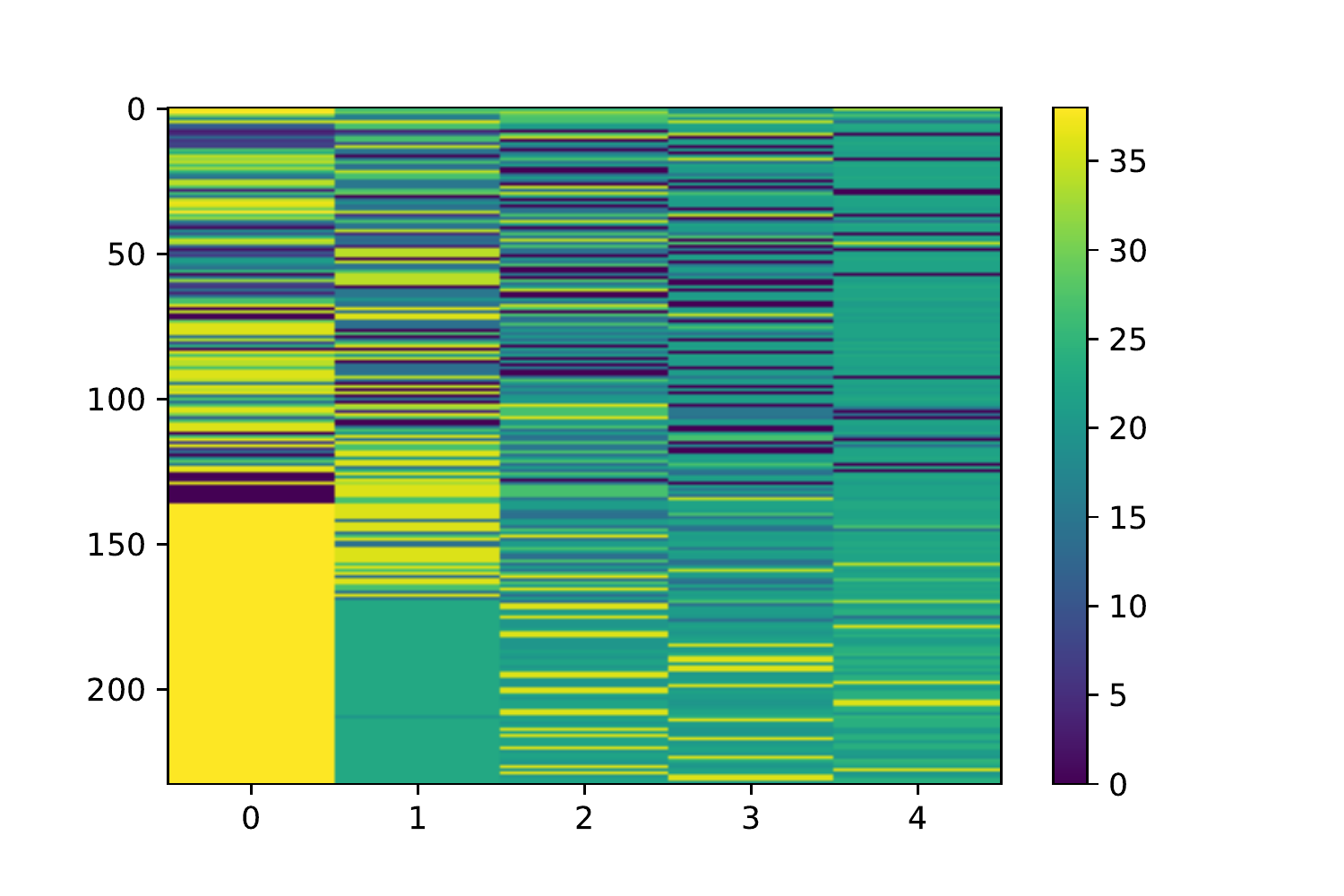}
\caption{Sampling Pattern}
\label{covid_sampling}
\end{figure}

\begin{figure}
    \centering
    \subfigure[Sampling frequency during in control phase]{\includegraphics[width=0.47\textwidth]{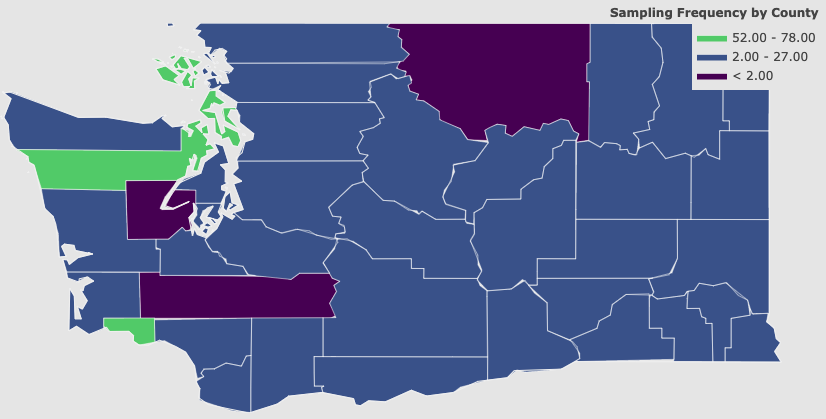}\label{in_control_sample}}
    \subfigure[Sampling frequency during out control phase]{\includegraphics[width=0.47\textwidth]{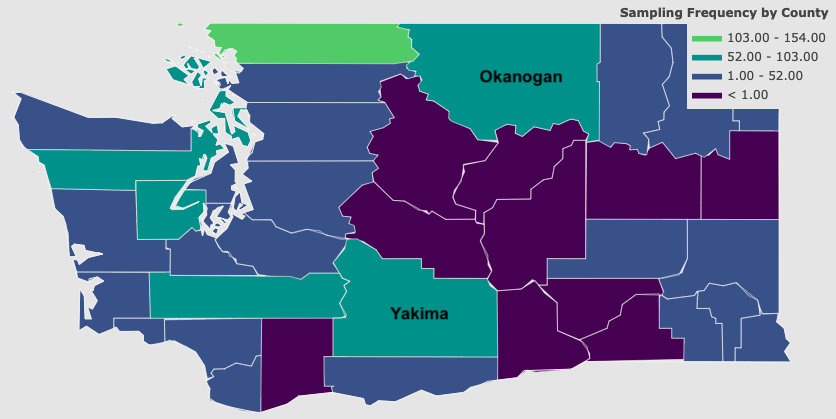}\label{out_control_sample}}
    \caption{Sampling Pattern for Covid-19 Case}\label{sample_covid}
\end{figure}

In the case study, we will use the real COVID-19 test report data from Johns Hopkins University Center for Systems Science and Engineering (JHU CSSE) (\citet{dong2020interactive}). The dataset is available on \url{https://github.com/CSSEGISandData/COVID-19}. 
More specifically, we will use the confirmed COVID-19 cases from all 39 counties in Washington state. 
 The source of daily positive cases in Washington State is the Department of Health (\url{https://www.doh.wa.gov/Emergencies/COVID19}). The time-series data is updated daily around 23:59 (UTC). We will use the data from Jan 23, 2020, to Sep 13, 2020, a total of 235 days, as an illustration. On each day, the confirmed cases are recorded in all counties in the United States. Overall, in the case study, we assume at each time epoch, the state government will focus on doing screening tests on Yakima County and Okanogan County over the $39$ counties. In this example, it is important to detect the outbreak in each individual county. Therefore, we set the failure mode as the mean shift of the infection rate of each individual county. The outbreak time for Yakima and Okanogan are 118 and 169, as shown in Table. \ref{covid_case}. The proposed method is able to detect the outbreak rapidly. TSSRP method can also detect the outbreak with some delay, while TRAS method cannot detect the outbreak. We further present the sampling pattern shown in Fig.~\ref{covid_sampling} and Fig.~\ref{sample_covid}. As shown in Fig.~\ref{covid_sampling}, the algorithm started to focus on certain counties starting around 140. Fig.~\ref{sample_covid} presents the aggregated sampling frequency for each county. It can be seen that the algorithm samples all the counties evenly during the in-control phase, as shown in Fig.~\ref{in_control_sample}. During the out-of-control phase, as shown in Fig.~\ref{out_control_sample}, it started to focus on the true hotspot counties such as Yakima and Okanogan, which helps us identify the outbreak quickly in those counties.

\begin{figure}
    \centering
    \subfigure[Monitoring Statistics for Hotspot Detection]{\includegraphics[width=0.47\textwidth]{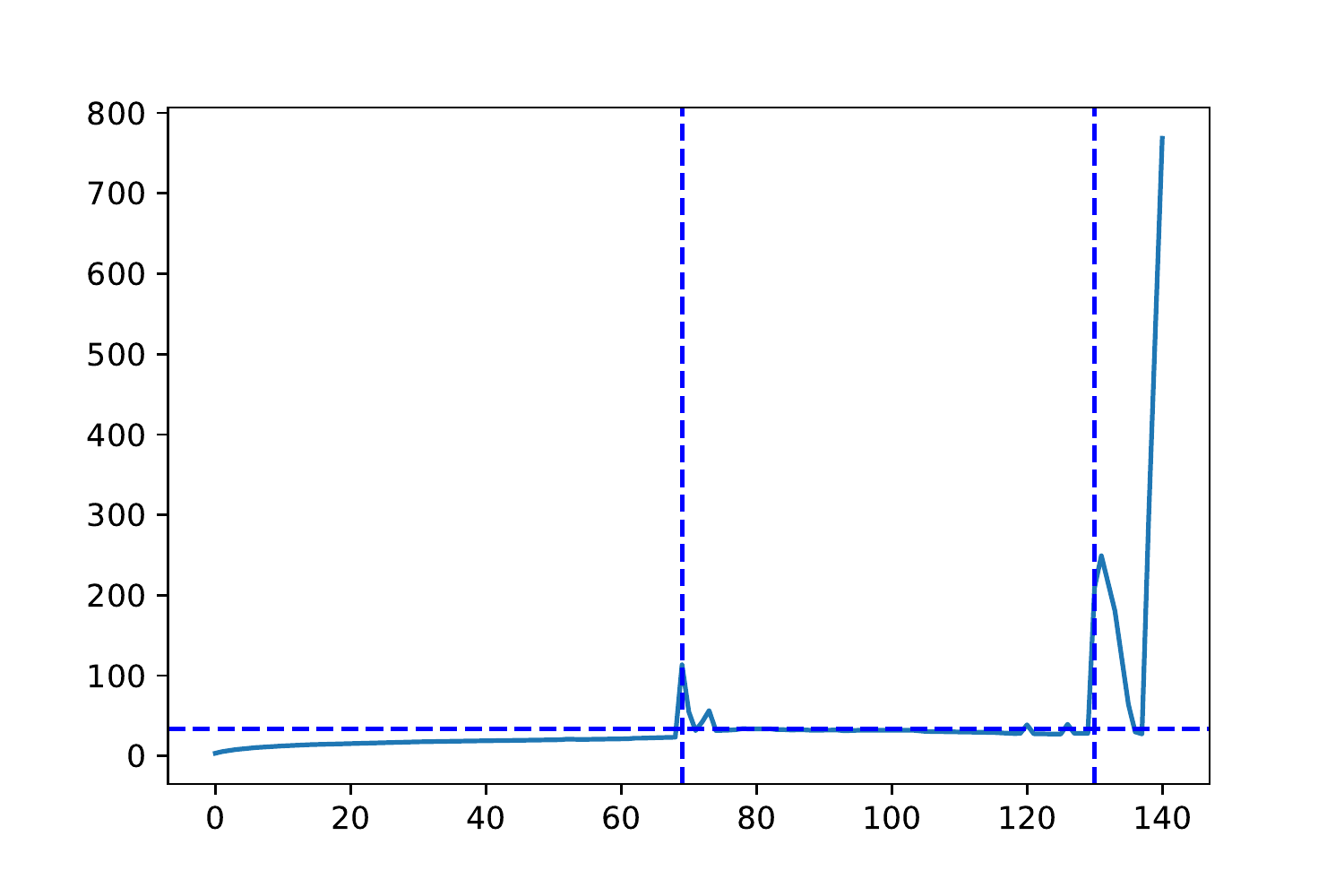}\label{covid_statistics}}
    \subfigure[OMonitoring Statistics for different counties]{\includegraphics[width=0.47\textwidth]{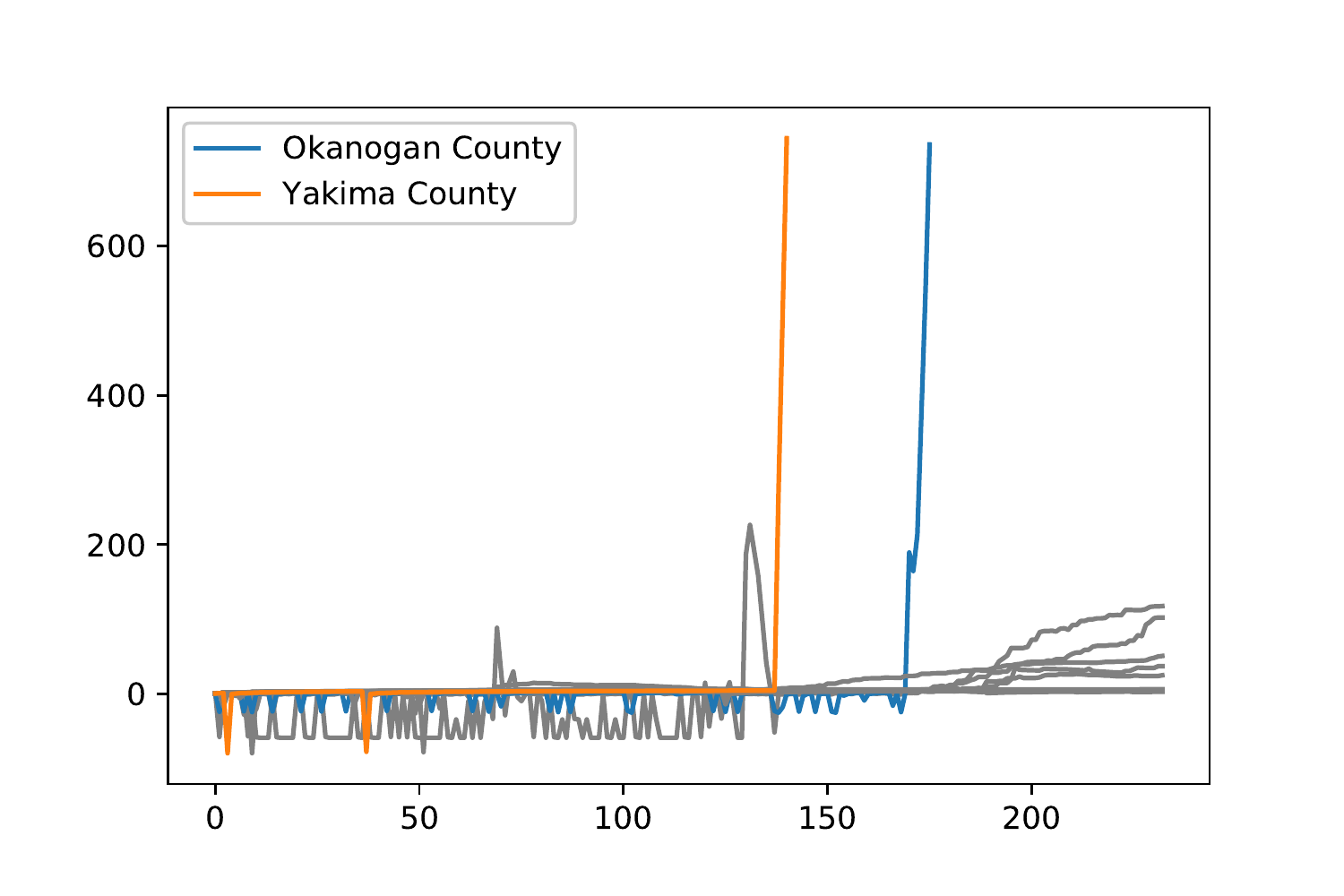}\label{covid_statistics_failure}}
    \caption{Monitoring Statistics for Covid-19 Case}\label{covid}
\end{figure}

\begin{table}[h!]
\caption{Detection Time for Covid hotspot detection}

\centering %
\begin{tabular}{|cc|cc|}
\hline
\multicolumn{2}{|c|}{} & \multicolumn{2}{c|}{Time of first signal} \\ \hline
\multicolumn{2}{|c|}{Location} & \multicolumn{1}{c|}{\begin{tabular}[c]{@{}c@{}}Yakima County\end{tabular}} & \begin{tabular}[c]{@{}c@{}}Okanogan County\end{tabular} \\ \hline
\multicolumn{2}{|c|}{Infection Rate $>$ 0.01} & \multicolumn{1}{c|}{118} & 169 \\ \hline
\multicolumn{1}{|c|}{\multirow{3}{*}{\begin{tabular}[c]{@{}c@{}}Competing \\ Methods\end{tabular}}} & MTSSRP & \multicolumn{1}{c|}{138} & 170 \\ \cline{2-4} 
\multicolumn{1}{|c|}{} & TSSRP & \multicolumn{1}{c|}{173} & 183 \\ \cline{2-4} 
\multicolumn{1}{|c|}{} & TRAS & \multicolumn{1}{c|}{-} & - \\ \hline
\end{tabular}\label{covid_case}
\end{table}

\section{Conclusion} \label{sec: conslusion}
Online change detection of high-dimensional data under multiple failure modes is an important problem in reality. In this paper, we propose to borrow the concept from Bayesian change point detection and MAB to adaptively sample useful local components, given the distributions of the multiple failure modes. Our proposed algorithm can balance between exploration of all possible system failure modes or exploitation of the most probable system failure mode. Furthermore, we also studied the properties of the proposed methods and showed the proposed algorithm could isolate the correct failure mode. Our simulation and case study show that the proposed algorithm, by considering the failure mode information, can significantly reduce the detection delay. 

\appendix

\section{Proof of Proposition \ref{thm: sort}}
\label{proofsort}
\begin{proof}
It is worth noting that under the assumption that distribution is independent, $\tilde{r}_{k,t}$ can be derived as
 
 \begin{align*}
 \tilde{S}_{t} & =\sum_{k=1}^{K_{s}} \tilde{r}_{(k),t} \\
 &= \sum_{k=1}^{K_{s}} ( \log(\exp(r_{(k),t-1})+1)+\sum_{j \in C_t} \log \frac{f_{j, (k)}(\tilde{x}^k_{j, t})}{f_{j,0}(\tilde{x}^k_{j, t})}) \\
 & = \sum_{k=1}^{K_{s}} ( \log(\exp(r_{(k),t-1})+1)) +  \sum_{j \in C_t} \sum_{k=1}^{K_{s}} \log \frac{f_{j, (k)}(\tilde{x}^k_{j, t})}{f_{j,0}(\tilde{x}^k_{j, t})}) \\
 & = C_0 + \sum_{j \in C_t} s_j
\end{align*}

 Therefore, $\tilde{S}_{t}$ can be derived as $\tilde{S}_{t}  =\sum_{k=1}^{K_{s}} \tilde{r}_{(k),t}  = C_0 + \sum_{j \in C_t} s_j$.
 Here, $s_j  = \sum_{k=1}^{K_{s}} \log \frac{f_{j, (k)}(\tilde{x}^k_{j, t})}{f_{j,0}(\tilde{x}^k_{j, t})}$ and $C_0 =\sum_{k=1}^{K_{s}} ( \log(\exp(r_{k,t-1})+1)) $ is a con-stance. To optimize  $\tilde{S}_{t+1} $, we can always select the largest $q$ from $s_j$ after ranking $s_{(1)} \geq s_{(2)} \geq \cdots \geq s_{(q)} \geq s_{(q+1)} \geq \cdots \geq s_{(p)}$. 

\end{proof}
\vspace*{-10mm}

\section{Proof of Proposition \ref{GaussianCase}}
\label{proofGaussian}

Therefore, $\log \frac{f_{k,j}(X_{j, t})}{f_{0,j}(X_{j, t})} =(\frac{1}{\sigma_{k,j}^{2}}(\tilde{x}^k_{j, t}-\mu_{k,j})^{2}-\frac{1}{\sigma_{k,0}^{2}}(\tilde{x}^k_{j, t}-\mu_{k,j})^{2})$. Therefore, we know that 

$$s_j  = \sum_{k=1}^{K_{s}} \log \frac{f_{j, (k)}(\tilde{x}^k_{j, t})}{f_{j,0}(\tilde{x}^k_{j, t})} =  \sum_{k=1}^{K_{s}}  (\frac{1}{\sigma_{k,j}^{2}}(\tilde{x}^k_{j, t}-\mu_{k,j})^{2}-\frac{1}{\sigma_{k,0}^{2}}(\tilde{x}^k_{j, t}-\mu_{k,j})^{2}).$$

\section{Proof of Proposition \ref{iid}}
\label{proofiid}
\begin{proof}
Here, 
If sensor $k$ is not selected,  $\frac{{f}_{k}(y_{t})}{{f}_{0}(y_{t})} = 1$. Therefore,  $R_{k,t}=R_{k,t-1}+1$. This implies that if sensor $k$ is not observed, the corresponding $R_{k,t}$ will increase by $1$. Furthermore, if sensor $k$ is selected, we will update $R_{k,t}$ as  $R_{k,t}= \frac{g(\tilde{x}_{k,t})}{f(\tilde{x}_{k,t})} (R_{k,t-1}+1)$.
\end{proof}
\section{Proof of Theorem \ref{Run}}
    \begin{proof}
    Let $T_1=\inf\{t\geq 1:  R_{(1),t} \geq A_1  \}$, $T_2=\inf\{t\geq 1: r_{(1),t} \geq A_2 \}$. 
    Since the event $\{R_{(1),t}\ge A_1\}$ is equivalent to the event $\{r_{(1),t}\ge \log A_1\}$, we only need to consider the behavior for $T_1$.
    
    For an upper bound of $T_1$, we define $T'=\inf\{t:R_{k,t}\ge A_1\}$ for some $k$. The theorem from \cite{pollak1987average} is valid in multivariate cases and thus $\mathbb E T_1\le \mathbb E T'=O(A_1)$.
    We show that $\sum_{k=1}^{K} R_{k,t}-K t$ is a martingale under the null hypothesis.
\begin{align*}
    \mathbb E [\sum_{k=1}^{K} R_{k,t+1}-K (t+1)|\mathcal{F}_t]&=\mathbb E\left[\sum_{k=1}^{K}(R_{k,t}+1)\frac{\tilde{f}_{C_{t+1}, k}({\bf X}_{C_{t+1}, t
+1})}{\tilde{f}_{C_{t+1},0}({\bf X}_{C_{t+1}, t+1})}-K(t+1)|\mathcal F_t\right]\\
&=\sum_{k=1}^{K}(R_{k,t}+1)\mathbb E\left[\frac{\tilde{f}_{C_{t+1}, k}({\bf X}_{C_{t+1}, t
+1})}{\tilde{f}_{C_{t+1},0}({\bf X}_{C_{t+1}, t+1})}|\mathcal F_t\right]-K(t+1)
\end{align*}
For any $k$, we have that 
\begin{align*}
    E\left[\frac{\tilde{f}_{C_{t+1}, k}({\bf X}_{C_{t+1}, t
+1})}{\tilde{f}_{C_{t+1},0}({\bf X}_{C_{t+1}, t+1})}|\mathcal F_t\right]&=
\int_{x_1,\ldots,x_n} \frac{\tilde{f}_{C_{t+1}, k}({\bf X}_{C_{t+1}, t
+1})}{\tilde{f}_{C_{t+1},0}({\bf X}_{C_{t+1}, t+1}) } f_0 (x_1,\ldots,x_n)\\
&=\int_{x_1,\ldots,x_n} \tilde{f}_{C_{t+1}, k}({\bf X}_{C_{t+1}, t
+1}) f_{C_{t+1},0}(x_{unobserved}|{\bf X}_{C_{t+1},t+1})\\
&=\int_{x_{observed}} \tilde{f}_{C_{t+1}, k}({\bf X}_{C_{t+1}, t
+1}) \int_{x_{unobserved}}f_{C_{t+1},0}(x_{unobserved}|{\bf X}_{C_{t+1},t+1})\\
&=1
\end{align*}
Thus, $ \mathbb E [\sum_{k=1}^{K} R_{k,t+1}-K (t+1)|\mathcal{F}_t]= \sum_{k=1}^{K} R_{k,t}-K t$ and $\sum_{k=1}^{K} R_{k,t}-K t$ is a martingale. Since we have that $\liminf_{t\to\infty} \int_{T_1>t}|\sum_{k=1}^{K} R_{k,t}-K t|{\rm d}\mathbb P_\infty=0$, $\sum_{k=1}^{K} R_{k,t}-K t$ is uniformly integrable. Thus $\mathbb E_\infty (\sum_{k=1}^{K} R_{k,T_1}-K T_1)=\mathbb E_\infty (\sum_{k=1}^{K} R_{k,0})=0$. Thus, $\mathbb E K T_1=\mathbb{E}_\infty \sum_{k=1}^{K} R_{k,T_1}\ge A_1$. \\
This completes the part for $T_1$. When $A_2=\log A_1$, the events  $\{R_{(1),t}\ge A_1\}$ and $\{r_{(1),t}\ge \log A_1\}$ are equivalent. Therefore, the behavior of $T_2$ follows the same rule. 

    \end{proof}
\section{Proof of Theorem \ref{Limit}}\label{proof-limit}
    \begin{proof}
   From \citep{pollak1987average}, we have
$$
\log R_{k,\nu+r}=\sum_{t=\nu+1}^{\nu+r} Z_{k,t}+\log R_{k,\nu}+\sum_{i=0}^{r-1}\log[1+\frac{1}{R_{k,\nu+i}}]
$$
The increment of $R_{k,t}$ will be almost $Z_{k,t}$ where $Z_{k,t}=\log(\frac{\tilde{f}_{C_{t}, k}({\bf X}_{C_{t}, t
})}{\tilde{f}_{C_{t},0}({\bf X}_{C_{t}, t})})$.
For any $l\neq k$,
$$
    \lim_{t\to\infty}\mathbb P(R_{l,t}\le R_{k,t}) = \lim_{T\to\infty} \mathbb P(\sum_{t=\nu+1}^T Z_{l,t}-Z_{k,t}\le b),
$$
where $b$ is some constant.
\begin{align*}
    \mathbb P(\sum_{t=\nu+1}^T Z_{l,t}-Z_{k,t}\le b) &=\mathbb P\left(\sum_{t=\nu+1}^T \log(\frac{\tilde{f}_{C_{t}, l}({\bf X}_{C_{t}, t})}{\tilde{f}_{C_{t},k}({\bf X}_{C_{t}, t})})\le b\right)
\end{align*}
${\bf X}_{C_t}$ is a sample of $q$ variables from $p$ dimensional data $\mathbf{{\bf X}}_t$. There are $\binom{p}{q}$ kinds of different samples. We divide the variables $\log(\frac{\tilde{f}_{C_{t}, l}({\bf X}_{C_{t}, t})}{\tilde{f}_{C_{t},k}({\bf X}_{C_{t}, t})})$ by the categories of different sample results, and denote them by $Y_{j,t}=\log(\frac{\tilde{f}_{C_{t}, l}({\bf X}_{C_{t}, t})}{\tilde{f}_{C_{t},k}({\bf X}_{C_{t}, t})})$, $j=1,2,\ldots,\binom{p}{q}$ and different $j$ represents different selection of $C_t$. We notice that when $j$ or $t$ is different, it has to be sampled from different times. Thus, they are independent.
 Since we consider the situation when $T$ goes to infinity, there exists at least one selection that is observed infinitely many times. 
 We first consider the case that there is only one selection observed infinitely many times. Without loss of generality, let this selection be the first one. 
 Let the times that other selections of data is observed be $c_2,\ldots,c_{\binom{p}{q}}$ respectively. 
 Denote $Y_{j,t}=\log(\frac{f_{l,j}(X_{t,j})}{f_{k,j}(X_{t,j})})$. When $j=1$, $t=1,2,\ldots$; when $j=2,\ldots,\binom{p}{q}$, $t=1,2,\ldots,c_{\binom{p}{q}}$. 
 It suffices to show $\mathbb P\left(\sum_{t=1}^\infty Y_{1,t}+\sum_{t=1}^{c_2}Y_{2,t}+\cdots+\sum_{t=1}^{c_{\binom{p}{q}}}Y_{\binom{p}{q},t}\le b\right)=1$ for any constant $b$. 
 Since $Y_{1,t}$ is i.i.d with negative expectation, we have that for any constant $b$, $\mathbb P\left(\sum_{t=1}^\infty Y_{1,t}\leq b\right)=1$\\
For any $\varepsilon>0$, let $d_j$ be the constant s.t. $\mathbb P\left(\sum_{t=1}^{c_j}Y_{j,t}<d_j\right)>1-\frac{\varepsilon}{\binom{p}{q}-1}$.\\
\begin{align*}
    &\mathbb{P}\left(\sum_{t=1}^{\infty}Y_{1,t}+\sum_{t=1}^{c_{2}}Y_{2,t}+\cdots+\sum_{t=1}^{c_{\binom{p}{q}}}Y_{\binom{p}{q},t}\le b\right) \\
    &>\mathbb{P}(\sum_{t=1}^{c_{2}}Y_{2,t}<d_2)\mathbb{P}(\sum_{t=1}^{c_{3}}Y_{3,t}<d_3)\cdots\mathbb{P}(\sum_{t=1}^{c_{\binom{p}{q}}}Y_{{\binom{p}{q}},t}<d_{\binom{p}{q}})\mathbb{P}\left(\sum_{t=1}^{\infty}Y_{1,t}\le b-d_{2}-\ldots-d_{\binom{p}{q}}\right)\\
    &>(1-\frac{\varepsilon}{\binom{p}{q}-1})^{\binom{p}{q}-1}\mathbb{P}\left(\sum_{t=1}^{\infty}Y_{1,t}\le b-d_{2}-\ldots-d_{\binom{p}{q}}\right)>1-\varepsilon
\end{align*}
Notice that this is true for any $\varepsilon>0$. We have $$\mathbb P\left(\sum_{t=1}^\infty Y_{1,t}+\sum_{t=1}^{c_2}Y_{2,t}+\cdots+\sum_{t=1}^{c_{\binom{p}{q}}}Y_{{\binom{p}{q}},t}\le b\right)=1.$$ Using similar method, we can also prove the case when there are more than one dimension observed infinitely many times. Thus, $\lim_{t\to\infty}\mathbb P(R_{l,t}\le R_{k,t})=1$. Since $l$ is chosen arbitrarily, we complete the proof.
    \end{proof}    
    
\section{Proof of Corollary \ref{limit-cor}}

\begin{proof}
We extend the notation from Appendix \ref{proof-limit}. For any $k\in\mathcal K$ and $l \notin \mathcal K$, consider
\begin{align*}
    \mathbb P(\sum_{t=\nu+1}^T Z_{l,t}-Z_{k,t}\le b) &=\mathbb P\left(\sum_{t=\nu+1}^T \log(\frac{\tilde{f}_{C_{t}, l}({\bf X}_{C_{t}, t})}{\tilde{f}_{C_{t},k}({\bf X}_{C_{t}, t})})\le b\right)
\end{align*}
\end{proof}
Since $k\in\mathcal K$, there exists some $t_0$ s.t. ${\bf X}_t\sim f_k$. Since the support of $f_k$ and $f_l$ are different, $\log(\frac{\tilde{f}_{C_{t_0}, l}({\bf X}_{C_{t_0}, t_0})}{\tilde{f}_{C_{t_0},k}({\bf X}_{C_{t_0}, t_0})})=-\infty$. For other times that ${\bf X}_t\nsim f_k$, we have an undefined term of $\log\frac 0 0$. Since this term can be interpreted as the log-likelihood difference at time $t$ and at this time, neither $l$ and $k$ are selected, we may assume $\log \frac 0 0=0$ under this assumption. Therefore,
$$
 \mathbb P(\sum_{t=\nu+1}^T Z_{l,t}-Z_{k,t}\le b)=1
$$

\bibliographystyle{plainnat}
\bibliography{ref.bib}

\begin{thebibliography}{39}
\providecommand{\natexlab}[1]{#1}
\providecommand{\url}[1]{\texttt{#1}}
\expandafter\ifx\csname urlstyle\endcsname\relax
  \providecommand{\doi}[1]{doi: #1}\else
  \providecommand{\doi}{doi: \begingroup \urlstyle{rm}\Url}\fi

\bibitem[CDC(2022)]{covid192022}
CDC.
\newblock Overview of testing for sars-cov-2 (covid-19), 2022.
\newblock URL
  \url{https://www.cdc.gov/coronavirus/2019-ncov/hcp/testing-overview.html}.

\bibitem[Chan et~al.(2017)]{chan2017optimal}
Hock~Peng Chan et~al.
\newblock Optimal sequential detection in multi-stream data.
\newblock \emph{The Annals of Statistics}, 45\penalty0 (6):\penalty0
  2736--2763, 2017.

\bibitem[Chang and Yadama(2010)]{chang2010statistical}
Shing~I. Chang and Srikanth Yadama.
\newblock Statistical process control for monitoring non-linear profiles using
  wavelet filtering and b-spline approximation.
\newblock \emph{International Journal of Production Research}, 48\penalty0
  (4):\penalty0 1049--1068, 2010.

\bibitem[Chen et~al.(2020)Chen, Zhang, and Poor]{chen2020bayesian}
Jie Chen, Wenyi Zhang, and H~Vincent Poor.
\newblock A bayesian approach to sequential change detection and isolation
  problems.
\newblock \emph{IEEE Transactions on Information Theory}, 67\penalty0
  (3):\penalty0 1796--1803, 2020.

\bibitem[Cho and Fryzlewicz(2015)]{cho2015multiple}
Haeran Cho and Piotr Fryzlewicz.
\newblock Multiple-change-point detection for high dimensional time series via
  sparsified binary segmentation.
\newblock \emph{Journal of the Royal Statistical Society: Series B: Statistical
  Methodology}, pages 475--507, 2015.

\bibitem[Colosimo and Grasso(2018)]{colosimo2018spatially}
Bianca~M Colosimo and Marco Grasso.
\newblock Spatially weighted pca for monitoring video image data with
  application to additive manufacturing.
\newblock \emph{Journal of Quality Technology}, 50\penalty0 (4):\penalty0
  391--417, 2018.

\bibitem[Dong et~al.(2020)Dong, Du, and Gardner]{dong2020interactive}
Ensheng Dong, Hongru Du, and Lauren Gardner.
\newblock An interactive web-based dashboard to track covid-19 in real time.
\newblock \emph{The Lancet infectious diseases}, 20\penalty0 (5):\penalty0
  533--534, 2020.

\bibitem[G{\'o}mez et~al.(2022)G{\'o}mez, Li, and Paynabar]{gomez2022adaptive}
Ana Mar{\'\i}a~Estrada G{\'o}mez, Dan Li, and Kamran Paynabar.
\newblock An adaptive sampling strategy for online monitoring and diagnosis of
  high-dimensional streaming data.
\newblock \emph{Technometrics}, 64\penalty0 (2):\penalty0 253--269, 2022.

\bibitem[Gopalan et~al.(2021)Gopalan, Lakshminarayanan, and
  Saligrama]{gopalan2021bandit}
Aditya Gopalan, Braghadeesh Lakshminarayanan, and Venkatesh Saligrama.
\newblock Bandit quickest changepoint detection.
\newblock \emph{Advances in Neural Information Processing Systems},
  34:\penalty0 29064--29073, 2021.

\bibitem[Grasso et~al.(2014)Grasso, Colosimo, and Pacella]{grasso2014profile}
M~Grasso, BM~Colosimo, and M~Pacella.
\newblock Profile monitoring via sensor fusion: the use of pca methods for
  multi-channel data.
\newblock \emph{International Journal of Production Research}, 52\penalty0
  (20):\penalty0 6110--6135, 2014.

\bibitem[Grasso et~al.(2017)Grasso, Laguzza, Semeraro, and
  Colosimo]{grasso2017process}
Marco Grasso, Vittorio Laguzza, Quirico Semeraro, and Bianca~Maria Colosimo.
\newblock In-process monitoring of selective laser melting: spatial detection
  of defects via image data analysis.
\newblock \emph{Journal of Manufacturing Science and Engineering}, 139\penalty0
  (5), 2017.

\bibitem[Li and Jin(2010)]{liOptimal2010}
Jing Li and Jionghua Jin.
\newblock Optimal sensor allocation by integrating causal models and
  set-covering algorithms.
\newblock \emph{IIE Transactions}, 42\penalty0 (8):\penalty0 564--576, May
  2010.

\bibitem[Liu and Shi(2013)]{liuObjectiveoriented2013}
Kaibo Liu and Jianjun Shi.
\newblock Objective-oriented optimal sensor allocation strategy for process
  monitoring and diagnosis by multivariate analysis in a {{Bayesian}} network.
\newblock \emph{IIE Transactions}, 45\penalty0 (6):\penalty0 630--643, June
  2013.

\bibitem[Liu et~al.(2013)Liu, Zhang, and Shi]{liu2013adaptive}
Kaibo Liu, Xi~Zhang, and Jianjun Shi.
\newblock Adaptive sensor allocation strategy for process monitoring and
  diagnosis in a bayesian network.
\newblock \emph{IEEE Transactions on Automation Science and Engineering},
  11\penalty0 (2):\penalty0 452--462, 2013.

\bibitem[Liu et~al.(2015)Liu, Mei, and Shi]{liuAdaptive2015}
Kaibo Liu, Yajun Mei, and Jianjun Shi.
\newblock An {{Adaptive Sampling Strategy}} for {{Online High}}-{{Dimensional
  Process Monitoring}}.
\newblock \emph{Technometrics}, 57\penalty0 (3):\penalty0 305--319, July 2015.

\bibitem[Lorden(1971)]{lorden1971procedures}
Gary Lorden.
\newblock Procedures for reacting to a change in distribution.
\newblock \emph{The Annals of Mathematical Statistics}, pages 1897--1908, 1971.

\bibitem[Malladi and Speyer(1999)]{malladi1999generalized}
Durga~P Malladi and Jason~L Speyer.
\newblock A generalized shiryayev sequential probability ratio test for change
  detection and isolation.
\newblock \emph{IEEE Transactions on Automatic Control}, 44\penalty0
  (8):\penalty0 1522--1534, 1999.

\bibitem[Mei(2010)]{mei2010efficient}
Yajun Mei.
\newblock Efficient scalable schemes for monitoring a large number of data
  streams.
\newblock \emph{Biometrika}, 97\penalty0 (2):\penalty0 419--433, 2010.

\bibitem[Mei(2011)]{mei2011quickest}
Yajun Mei.
\newblock Quickest detection in censoring sensor networks.
\newblock In \emph{2011 IEEE International Symposium on Information Theory
  Proceedings}, pages 2148--2152. IEEE, 2011.

\bibitem[Nikiforov et~al.(1993)Nikiforov, Varavva, and
  Kireichikov]{nikiforov1993application}
I~Nikiforov, V~Varavva, and V~Kireichikov.
\newblock Application of statistical fault detection algorithms to navigation
  systems monitoring.
\newblock \emph{Automatica}, 29\penalty0 (5):\penalty0 1275--1290, 1993.

\bibitem[Nikiforov(1995)]{nikiforov1995generalized}
Igor~V Nikiforov.
\newblock A generalized change detection problem.
\newblock \emph{IEEE Transactions on Information theory}, 41\penalty0
  (1):\penalty0 171--187, 1995.

\bibitem[Paynabar and Jin(2011)]{paynabar2011characterization}
Kamran Paynabar and Jionghua Jin.
\newblock Characterization of non-linear profiles variations using mixed-effect
  models and wavelets.
\newblock \emph{IIE transactions}, 43\penalty0 (4):\penalty0 275--290, 2011.

\bibitem[Paynabar et~al.(2016)Paynabar, Zou, and Qiu]{paynabar2015change}
Kamran Paynabar, Changliang Zou, and Peihua Qiu.
\newblock A change-point approach for phase-i analysis in multivariate profile
  monitoring and diagnosis.
\newblock \emph{Technometrics}, 58\penalty0 (2):\penalty0 191--204, 2016.

\bibitem[Pollak(1985)]{pollak1985optimal}
Moshe Pollak.
\newblock Optimal detection of a change in distribution.
\newblock \emph{The Annals of Statistics}, pages 206--227, 1985.

\bibitem[Pollak(1987)]{pollak1987average}
Moshe Pollak.
\newblock Average run lengths of an optimal method of detecting a change in
  distribution.
\newblock \emph{The Annals of Statistics}, pages 749--779, 1987.

\bibitem[Qiu et~al.(2010)Qiu, Zou, and Wang]{qiu2010nonparametric}
Peihua Qiu, Changliang Zou, and Zhaojun Wang.
\newblock Nonparametric profile monitoring by mixed effects modeling.
\newblock \emph{Technometrics}, 52\penalty0 (3), 2010.

\bibitem[Ren et~al.(2020)Ren, Zou, Chen, and Li]{ren2020large}
Haojie Ren, Changliang Zou, Nan Chen, and Runze Li.
\newblock Large-scale datastreams surveillance via pattern-oriented-sampling.
\newblock \emph{Journal of the American Statistical Association}, pages 1--15,
  2020.

\bibitem[Shiryaev(1963)]{shiryaev1963optimum}
Albert~N Shiryaev.
\newblock On optimum methods in quickest detection problems.
\newblock \emph{Theory of Probability \& Its Applications}, 8\penalty0
  (1):\penalty0 22--46, 1963.

\bibitem[Wang et~al.(2018)Wang, Xian, Tsung, and Liu]{wangspatialadaptive2018}
Andi Wang, Xiaochen Xian, Fugee Tsung, and Kaibo Liu.
\newblock A spatial-adaptive sampling procedure for online monitoring of big
  data streams.
\newblock \emph{Journal of Quality Technology}, 50\penalty0 (4):\penalty0
  329--343, October 2018.

\bibitem[Wang and Mei(2015)]{wang2015large}
Yuan Wang and Yajun Mei.
\newblock Large-scale multi-stream quickest change detection via shrinkage
  post-change estimation.
\newblock \emph{IEEE Transactions on Information Theory}, 61\penalty0
  (12):\penalty0 6926--6938, 2015.

\bibitem[Willsky(1976)]{willsky1976survey}
Alan~S Willsky.
\newblock A survey of design methods for failure detection in dynamic systems.
\newblock \emph{Automatica}, 12\penalty0 (6):\penalty0 601--611, 1976.

\bibitem[Xian et~al.(2018)Xian, Wang, and Liu]{xianNonparametric2018}
Xiaochen Xian, Andi Wang, and Kaibo Liu.
\newblock A {{Nonparametric Adaptive Sampling Strategy}} for {{Online
  Monitoring}} of {{Big Data Streams}}.
\newblock \emph{Technometrics}, 60\penalty0 (1):\penalty0 14--25, January 2018.

\bibitem[Xie and Siegmund(2013)]{xie2013sequential}
Yao Xie and David Siegmund.
\newblock Sequential multi-sensor change-point detection.
\newblock In \emph{2013 Information Theory and Applications Workshop (ITA)},
  pages 1--20. IEEE, 2013.

\bibitem[{Yan} et~al.(2015){Yan}, {Paynabar}, and {Shi}]{yan2015image}
H.~{Yan}, K.~{Paynabar}, and J.~{Shi}.
\newblock Image-based process monitoring using low-rank tensor decomposition.
\newblock \emph{IEEE Transactions on Automation Science and Engineering},
  12\penalty0 (1):\penalty0 216--227, 2015.

\bibitem[Yan et~al.(2017)Yan, Paynabar, and Shi]{yan2017anomaly}
Hao Yan, Kamran Paynabar, and Jianjun Shi.
\newblock Anomaly detection in images with smooth background via smooth-sparse
  decomposition.
\newblock \emph{Technometrics}, 59\penalty0 (1):\penalty0 102--114, 2017.

\bibitem[Yan et~al.(2018)Yan, Paynabar, and Shi]{yanRealTime2018}
Hao Yan, Kamran Paynabar, and Jianjun Shi.
\newblock Real-{{Time Monitoring}} of {{High}}-{{Dimensional Functional Data
  Streams}} via {{Spatio}}-{{Temporal Smooth Sparse Decomposition}}.
\newblock \emph{Technometrics}, 60\penalty0 (2):\penalty0 181--197, April 2018.

\bibitem[Yan et~al.(2020)Yan, Grasso, Paynabar, and Colosimo]{yan2020real}
Hao Yan, Marco Grasso, Kamran Paynabar, and Bianca~Maria Colosimo.
\newblock Real-time detection of clustered events in video-imaging data with
  applications to additive manufacturing.
\newblock \emph{arXiv preprint arXiv:2004.10977}, 2020.

\bibitem[Yue et~al.(2017)Yue, Yan, Park, Liang, and Shi]{yue2017wavelet}
Xiaowei Yue, Hao Yan, Jin~Gyu Park, Zhiyong Liang, and Jianjun Shi.
\newblock A wavelet-based penalized mixed-effects decomposition for
  multichannel profile detection of in-line raman spectroscopy.
\newblock \emph{IEEE Transactions on Automation Science and Engineering},
  15\penalty0 (3):\penalty0 1258--1271, 2017.

\bibitem[Zhang and Mei(2020)]{zhang2020bandit}
Wanrong Zhang and Yajun Mei.
\newblock Bandit change-point detection for real-time monitoring
  high-dimensional data under sampling control.
\newblock \emph{arXiv preprint arXiv:2009.11891}, 2020.

\end{thebibliography}
\end{document}